\documentclass[11pt]{article}
\usepackage[margin=1in]{geometry}
\usepackage[authoryear]{natbib}
\usepackage{graphicx}
\usepackage{booktabs}
\usepackage{subcaption}
\usepackage{enumitem}
\usepackage{amsmath}
\usepackage{float}
\usepackage[section]{placeins}
\usepackage[colorlinks=true,citecolor=blue,urlcolor=blue]{hyperref}
\usepackage{url}

\begin{document}

\title{TerraNostra-Bench: Benchmarking Spatial Downscaling and Temporal Rollout for Global 1-km Land-Use Reconstruction}

\author{
  Amirpasha Mozaffari$^{1,*}$ \and Marina Castaño$^{1}$ \and Stefano Materia$^{1,2}$ \and
  Joan Vedrí$^{1}$ \and Jordi Varela-Agrelo$^{3}$ \and Dario Garcia-Gasulla$^{3}$ \and
  Etienne Tourigny$^{1}$ \and Mario Acosta$^{1}$ \and Amanda Duarte$^{1}$
}

\date{\small
  $^1$Earth Science Department, Barcelona Supercomputing Center (BSC), Barcelona, Spain\\
  $^2$Foundation Euro-Mediterranean Center on Climate Change, Bologna, Italy\\
  $^3$AI Institute, Barcelona Supercomputing Center (BSC), Barcelona, Spain\\
  $^*$Corresponding author: amirpasha.mozaffari@bsc.es
}

\maketitle

\begin{abstract}
Uncertainty in the terrestrial carbon cycle remains a major constraint in climate projections, partly driven by imperfect land surface representations, and weaknesses in Earth system models. To address this limitation, we propose a data-driven framework for high-resolution reconstruction of land use, treating reconstruction as a semantic-segmentation task. The framework integrates coarse-resolution scenario data and climate classification with static geophysical features to produce annual land-use maps at 1~km resolution. Our contribution begins with the construction of a benchmark, TerraNostra-Bench. This change-aware evaluation protocol separates spatial downscaling from autoregressive temporal rollout and scores models on the rare land-use-change pixels rather than the dominant static background. Using this protocol we compare convolutional (U-Net) and transformer (Swin) backbones in autoregressive and non-\hspace{0pt}autoregressive schemes against persistence, linear, and shallow convolutional baselines. We find that the autoregressive prior is the most informative input yet accumulates error over long rollouts: autoregressive models lose 5.7--7.2 accuracy points over 13 years, against 2.2--2.8 for non-\hspace{0pt}autoregressive models. This trade-off is governed by autoregressive choice rather than the backbone. We then show that this trade-off can be tuned  by 1) an inference-time prior-trust threshold that requires no retraining and 2) the training-time weighting of the reconstruction and change objectives; the two levers trace complementary frontiers, with calibration recovering up to 10 accuracy points while retaining change sensitivity above the persistence floor. Finally, we evaluate extrapolation beyond the training window in both temporal directions-- backward to 1955--1959 and forward to 2015--2019 under scenario forcing---and find that the models retain most of their in-window skill and reproduce the same autoregressive trade-off, evidence that it is intrinsic to the design rather than to the test period. This work establishes the methodological foundation and evaluation surface for a forthcoming temporally continuous land-use product, with the aim of reducing critical uncertainties and improving the predictive power of next-generation climate simulations.

\end{abstract}

\section{Introduction}
The terrestrial carbon cycle is among the most uncertain components in climate modeling \citep{booth2012high}, reflecting both incomplete process understanding and limitations in Earth System Model (ESM) capabilities, particularly with respect to spatial resolution and parameterizations \citep{qiu2025uncertain}. Land-use and land-cover changes (LULCC) drive the largest variations in terrestrial carbon stocks, particularly when forests are converted to crops or pastures, which retain only a fraction of the carbon stored in forests \citep{houghton2018interactions}. Carbon emissions from LULCC suffer from many sources of uncertainty, including input data (both statistical-based and satellite-based), and the different methods used in models and inventories \citep{brasika2025uncertainties}.Indeed, relying on coarse or outdated land boundary data can critically misrepresent land--atmosphere exchanges. Studies have shown that coarse land cover spatial resolutions can introduce substantial biases in simulated terrestrial carbon sequestration, affecting its magnitude, interannual variability, and spatial distribution \citep{zhao2010spatial}. Given that land-use change has consistently driven substantial greenhouse-gas emissions over the last century \citep{friedlingstein2025global}, the accurate representation of high-resolution land boundary conditions is essential for reducing uncertainties in coupled carbon–water–energy fluxes \citep{gier2024representation}. The prospect to reduce the uncertainty in CO\textsubscript{2}-equivalent emissions from land use and land cover changes has been demonstrated by e.g. \cite{friedlingstein2022global}, who replaced the FAO statistics for land use with remote sensing data for Brazil.

\begin{figure}
 \centering
        \includegraphics[width=0.8\textwidth]{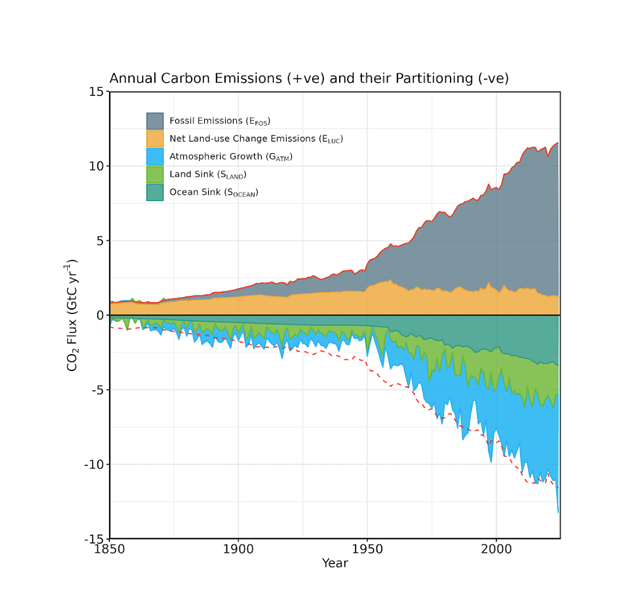}
 \caption{Annual global carbon emissions and their partitioning from 1850 to present. The data highlights the significant and persistent contribution of net land-use change emissions (ELUC, orange) to the atmospheric carbon burden, alongside fossil emissions. Source: Global Carbon Budget 2025 \citep{friedlingstein2025global} }
\label{fig1_lu_carbon}
\end{figure}

While several datasets offer valuable insights into land use, each falls short in terms of spatial resolution, temporal coverage (historical or future), or logical continuity \citep{hurtt2020harmonization, winkler2020hilda+,lv2025land,chen2020global}. To address this scaling mismatch, our work proposes a deep learning framework for reconstructing high-resolution land surface boundary conditions for ESMs. This architecture is designed as the foundation for a temporally continuous, spatially detailed land use dataset spanning historical, contemporary, and future periods in subsequent work. Specifically, our approach leverages static geophysical features and dynamic social and physical indicators to downscale the standard low-resolution land-use forcing typically used in ESMs, that is the Land Use Harmonization (LUH, \cite{hurtt2011harmonization}, and LUH2, \cite{hurtt2020harmonization}). Therefore, we aim at improving the representation of land surface heterogeneity that is necessary for enabling more accurate simulations of carbon, water, and energy fluxes over time.
A high-resolution land-surface forcing is necessary but not sufficient to reduce carbon, water and energy flux uncertainties in ESMs. Even when the surface is resolved at 1 km, granularity is far from perfect, biases are still present, and the simulated carbon, water and energy fluxes still depend on how each type is represented in the host land-surface scheme. However, this work represents an important step towards a LULCC constraint to the CMIP7 generation of ESMs, as an alternative to the LUH forcing datasets. Relying on an ensemble of land use forcings allows systematic sampling of LULCC uncertainty, and therefore to quantify potential biases and explore their impact on the land carbon sink.

\textbf{Scope and contributions.} This paper establishes the methodological framework and the benchmark dataset for training and assessing high-resolution land-use reconstruction models over the observation era covered by the HIstoric Land Dynamics Assessment (HILDA+, \cite{winkler2021global}). Our contributions are threefold. First, we frame high-resolution land-use reconstruction as a semantic-segmentation problem that fuses coarse dynamic forcings with high-resolution static priors. Second, we assemble a benchmark dataset with fixed spatial--temporal splits, and a suite of reference baselines. Its dual-task, change-aware evaluation protocol disentangles spatial downscaling from autoregressive temporal stability and foregrounds land-use \emph{change} rather than the dominant static background, enabling reproducible training and assessment of land-use downscaling models. Third, using this protocol we systematically compare autoregressive and non-\hspace{0pt}autoregressive variants of a convolutional (U-Net) and a transformer (Swin) backbone against persistence, linear, and shallow  convolutional baselines. Additionally, we analyse when conditioning on the previous year's map helps and when it hurts. The application of this framework to full historical and future inference (1850--2100) is the subject of follow-up work.

\section{Related Work}

The reconstruction and projection of land use have historically relied on rule-based harmonization \citep{gao2023clumondo}, such as the standard LUH2 framework \citep{hurtt2020harmonization}. 
In addition, methods like land-use translators are created to downscale historical and future changes to the regional scale \citep{hoffmann2022high}. However, a broad spectrum of ML techniques is increasingly employed to overcome the ``scaling mismatch'' of these coarse baselines and to generate high-resolution, spatially continuous datasets.

Pixel-wise and tree-based methods, particularly Random Forests (RF) and Extreme Gradient Boosting (XGBoost), have become a standard for high-resolution historical reconstruction owing to their ability to handle non-linear relationships in heterogeneous data, and have been widely adopted for diverse tasks, including cropland reconstruction \citep{yang2020integration} and land-use change estimation \citep{aboh2020satellite}. Similarly, the HILDA+ dataset employs data-driven fusion, utilizing RF to downscale coarse FAO statistics by learning relationships with fine-scale drivers like population density \citep{winkler2020hilda+}. XGBoost, another ensemble-based method, has shown strong performance in the historical reconstruction of LULCC and LAI \citep{mozaffari2025reconstruction}.

Dynamic Simulation and Hybrid Systems captures the temporal evolution and competition between land use classes. Historically, researchers have developed hybrid systems combining ML with Cellular Automata (CA) \citep{wolfram1983statistical}. The Future Land Use Simulation (FLUS) model integrates Artificial Neural Networks (ANNs) to estimate suitability probabilities, coupled with CA to simulate complex local competition and inertia \citep{xu2019simulation} \citep{chen2021integrating}. Building on systems theory, the Modified CLUMondo model employs RF to dynamically calculate ``competitive advantage'', allowing for the simulation of complex transitions in tipping-element regions under future climate pledges \citep{lv2025land}. Additionally, recurrent architectures like Long Short-Term Memory (LSTM) networks have been applied to model land use change by explicitly modelling temporal dependencies in historical time series \citep{zhao2023combining} \citep{yang202130}, and convolutional-recurrent variants such as ConvLSTM jointly capture spatial and temporal structure \citep{Shi2015}. A recurring difficulty for such temporally conditioned models is that, when rolled out autoregressively over many steps, prediction errors accumulate and can destabilise long-horizon trajectories---a trade-off we examine directly by contrasting autoregressive and non-\hspace{0pt}autoregressive variants. Convolutional Neural Networks (CNNs) and all their variations have emerged as a powerful tool for spatial downscaling and super-resolution, addressing the limitations of pixel-wise and rule-based methods. Unlike RF or CA, CNNs utilize spatial context to resolve texture and hierarchy, making them ideal for bridging the gap between coarse climatic forcing and kilometer-scale boundaries \citep{truong2024jaxa} \citep{zhu2015land} \citep{li2024learningexactguidanceupdating}, \citep{feng2025convolutional}. More recently, Vision Transformers and hybrid Transformer--U-Net designs such as Swin-Unet \citep{cao2021swinunetunetlikepuretransformer} extend this spatial modelling with self-attention, capturing long-range dependencies across large land formations that purely convolutional receptive fields may miss.

Each of these families addresses part of the downscaling problem while leaving a gap that our framework targets. Rule-based harmonization \citep{hurtt2020harmonization} preserves large-scale consistency but cannot synthesise fine-scale spatial texture; pixel-wise RF and XGBoost models capture non-linear driver relationships but discard spatial context; cellular-automata and FLUS-style hybrids rely on hand-tuned transition rules and are typically calibrated at regional scale; and recurrent temporal models capture sequence dynamics but not the joint spatial structure needed for global 1~km maps. Our framework itself was introduced for single-year reconstruction in \cite{mozaffari2026scalable}, where each year is predicted from an adjacent observed map supplied as a partially masked prior; the present work extends it to the multi-year setting, removing the prior entirely or replacing it with the model's own prediction in closed loop, with scheduled sampling replacing fixed masking during training. Because the two settings differ in the observational information available at inference, their absolute scores are not directly comparable---the single-year task is conditioned on a real adjacent observation, whereas the rollout receives none after its initial anchor. A further challenge  is the absence of a standardised evaluation surface: existing high-resolution land-use studies each adopt bespoke spatial splits, target taxonomies, and metrics, which precludes direct comparison and obscures whether learned models outperform trivial persistence baselines. While several remote-sensing benchmarks exist, they target fundamentally different settings: EuroSAT \citep{helber2019eurosat} and BigEarthNet \citep{sumbul2019bigearthnet} pose static, single-image land-cover classification; DynamicEarthNet \citep{toker2022dynamicearthnet} and PASTIS \citep{garnot2021pastis} provide short-horizon, regional satellite-image-time-series segmentation; and LandBench \citep{li2024landbench} targets continuous land-surface-variable prediction. None are designed for long-horizon, global land-use reconstruction with autoregressive temporal rollout and change-aware evaluation suitable as Earth-system-model boundary conditions, which require spatially complete and temporally consistent annual maps. We address this gap by introducing a fixed-split, change-aware benchmark with reference baselines, on which the models above are evaluated.

\section{Methods}
The methodology utilizes the data-driven framework described in  \cite{mozaffari2025historical,mozaffari2026scalable}, treating land-use reconstruction as a semantic segmentation task. We define a mapping $f_\theta: \mathcal{X} \to \mathcal{Y}$, where $\mathcal{X}$ represents coarse dynamic forcings and high-resolution static priors, and $\mathcal{Y}$ represents the 1~km land-use map.

\subsection{Data}
Our feature space integrates heterogeneous global datasets, including climatic, topographical, and anthropogenic variables, all resampled to a unified 1~km grid.

\textbf{Target Variable:} The \textbf{HILDA+} is a data-driven product, constructed by \citep{winkler2020hilda+} as a global land use at 1~km resolution by fusing FAO national statistics with fine-scale spatial drivers via Random Forest. We use HILDA+ as our target variable. The original scheme distinguishes 13 classes, including five forest sub-types separated by leaf type and phenology (evergreen/deciduous, needle-/broad-leaved). Because our coarse predictors carry no information that distinguishes these forest morpho-phenologies, we merge the five forest sub-types into a single \emph{Forest} class, yielding the seven land-use categories in Table~\ref{tab:lu_classes}. This aligns the target with the information actually present in the inputs, avoids penalizing the model for distinctions it cannot resolve, and mitigates the severe imbalance of the rare forest sub-types; an eighth label is added to denote the ocean. HILDA+ is a widely used reference for long-term land-use change: it combines multiple remote-sensing products, long-term land-use reconstructions and national statistics into an annual record from 1960 onwards, and its analysis showed that global land-use change is about four times larger than earlier estimates suggested \cite{winkler2021global}. It nevertheless remains a modelled product, so the metrics presented in this work quantify agreement with a modelled reference rather than with ground truth: they upper-bound agreement with HILDA+, not with the true land surface.

\textbf{Predictor Variables:} Predictors were chosen to cover the three groups of drivers that determine where each land-use class can occur: the coarse land-use trajectory itself, climate and human pressure, and static physical constraints (terrain and soil). The set follows our earlier single-year model \cite{mozaffari2026scalable} and is restricted to variables available globally and consistently for both the historical period and future scenarios, so the trained model can also be applied to future scenarios.
    
\begin{itemize}[leftmargin=*, topsep=2pt]
    \setlength\itemsep{0em}
    \item \textbf{Dynamic climatic \& anthropogenic forcing:} We employ the LUH2 dataset (historical and future) \citep{hurtt2020harmonization} at a coarse resolution of $0.25^{\circ} \times 0.25^{\circ}$ ($\sim$31~km), which supplies the coarse land-use trajectory that our model aims at downscaling: 12 fractional land-use state variables (each in $[0,1]$) together with two secondary-vegetation parameters (mean age and biomass). To stratify environmental conditions, we add Köppen-Geiger climate classification maps \citep{beck2023high}, which delimit the climatic envelope of viable biomes, and the ISIMIP2b population data \citep{frieler2017assessing} as a proxy for the anthropogenic pressure that drives urban and cropland expansion, both at 1~km resolution.
    \item \textbf{Static geophysical priors:} These items constrain the locations where each land-use class is physically plausible. We incorporate elevation and bathymetry from GEBCO 2024 \citep{GEBCO2024}, together with derived slope and aspect, which capture topographic suitability and the local microclimates that shape vegetation. Soil heterogeneity---clay, sand, and organic content from high-resolution Pacific Northwest National Laboratory (PNNL) global datasets \citep{li2024global}---encodes substrate fertility and drainage that constrain agriculture and natural cover. The source product provides each soil property on 6- and 10-layer vertical discretisations; we use the 6-layer version and collapse it into a single depth-weighted surface layer (Section~\ref{sec:preprocessing}), so that each soil property enters the model as one 2D field.
\end{itemize}

\begin{table}[htbp]
    \centering
    \caption{Overview of Land Use/Land Cover taxonomies and predictor variables. The table displays the original HILDA+ classes, the consolidated TerraNostra target classes, and the LUH2 predictor variable definitions.}
    \label{tab:lu_classes}
    \small 
    \setlength{\tabcolsep}{4pt} 
    \begin{tabular}{ll|ll|ll}
        \toprule
        \multicolumn{2}{c|}{\textbf{HILDA+ (Original)}} & \multicolumn{2}{c|}{\textbf{TerraNostra (Target)}} & \multicolumn{2}{c}{\textbf{LUH2 (Predictors)}} \\
        \textbf{Code} & \textbf{Class Name} & \textbf{Code} & \textbf{Class Name} & \textbf{Abbr.} & \textbf{Full Class Name} \\
        \midrule
        11 & Urban & 1 & Urban & primf & Primary Forest \\
        22 & Cropland & 2 & Cropland & primn & Primary Non-Forest \\
        33 & Pasture & 3 & Pasture & secdf & Secondary Forest \\
        40 & Forest (Unknown/Other) & 4 & Forest & secdn & Secondary Non-Forest \\
        41 & Forest (Evergreen, needle) & 5 & Grass/shrubland & pastr & Managed Pasture \\
        42 & Forest (Evergreen, broad) & 6 & Other land & range & Rangeland \\
        43 & Forest (Deciduous, needle) & 7 & Water & urban & Urban Land \\
        44 & Forest (Deciduous, broad) & & & c3ann & C3 Annual crops \\
        45 & Forest (Mixed) & & & c4ann & C4 Annual crops \\
        55 & Grass/shrubland & & & c3per & C3 Perennial crops \\
        66 & Other land & & & c4per & C4 Perennial crops \\
        77 & Water & & & c3nfx & C3 Nitrogen-fixing crops \\
        & & & & secma & Secondary Mean Age \\
        & & & & secmb & Secondary Mean Biomass \\
        \bottomrule
    \end{tabular}
\end{table}

\subsection{Preprocessing}\label{sec:preprocessing}
To ensure spatial and temporal consistency across diverse data sources, we established a standardized pipeline centered on Climate Data Operators (CDO) \citep{schulzweida2019cdo}. All predictor variables were aligned to the 1~km WGS84 grid and projection of the target HILDA+ dataset.

\textbf{Categorical \& Temporal Interpolation:} Köppen-Geiger climatic zones are defined as discrete variables available only at multi-decadal intervals. Therefore, for these units standard linear interpolation yields invalid intermediate classes, since the class codes are nominal labels and averaging them produces non-integer values or unrelated zones. We implemented a \textit{stochastic weighted dithering} algorithm, adapting the random-threshold principle of image dithering \citep{ulichney1987digital}, to bridge the 30-year gaps. We generated a static random noise field ($N \sim U[0,1]$) to serve as a spatially consistent probabilistic threshold.
For any intermediate year $t$, a pixel transitions from the start-period class to the end-period class if the noise value is lower than the normalized time weight $w(t)$. Equivalently, for each pixel whose climatic zone changes between two reference epochs, this assigns a single transition year within the 30-year window, drawn from a distribution whose cumulative probability grows linearly across the window. Once a pixel has transitioned to the new zone, it remains committed to it for the rest of the window. This procedure preserves the categorical nature of the variable while removing the staircase artefacts that a hard nearest-year assignment would introduce.

\textbf{Mass-Conserving Population Downscaling:} For ISIMIP2b population data that is at $0.5^{\circ} \times 0.5^{\circ}$ spatial resolution, direct interpolation of count data would violate mass conservation. We instead implemented a density-based downscaling workflow: raw population counts were converted to density ($people/m^2$) using source grid cell areas, conservatively remapped to the target grid, and multiplied by the target cell areas to recover local counts. Additionally, as the ISIMIP2b historical record begins in 1861, we extended the series back to 1850 using pixel-wise linear back-casting. Trends were derived from the initial decade (1861–1871), and extrapolated values were clamped at zero to prevent physical inconsistencies.

\textbf{Geophysical Feature Aggregation:} To generate surface-representative features from PNNL six-layer vertical soil data \citep{li2024global}, we computed a 2D non-linear depth-weighted average. We applied a decreasing weighting scheme of $w = [0.40, 0.25, 0.15, 0.10, 0.05, 0.05]$ to the vertical column, prioritizing the upper soil horizons where land–atmosphere interactions and biological activity are most pronounced.

\begin{figure}[t!]
  \centering
  \includegraphics[width=0.75\textwidth]{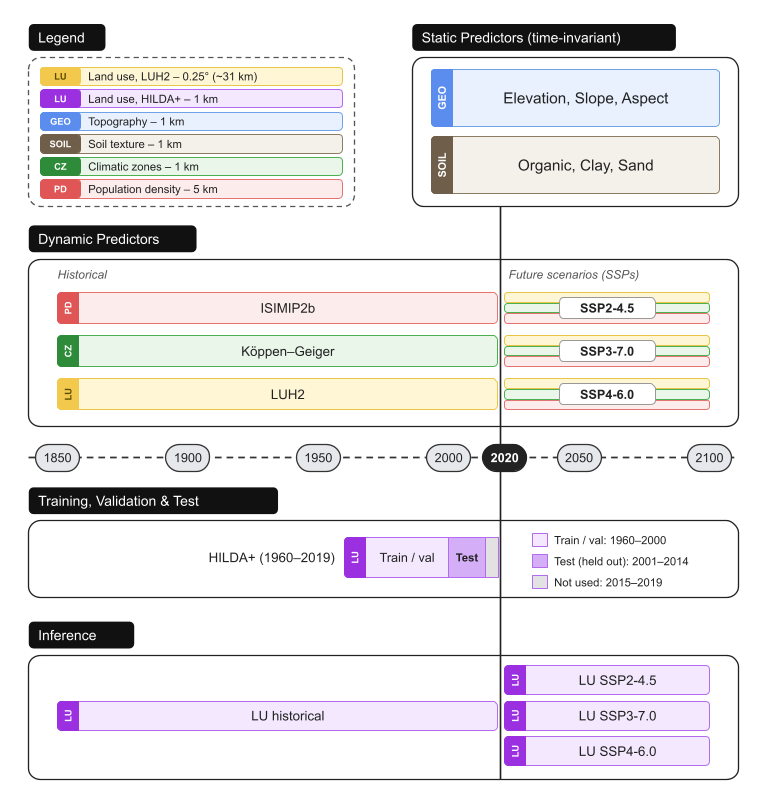}
  \caption{\textbf{Temporal Data Scope.} An overview of the dynamic and static predictors along with the target variable over the historical and future timelines.}
  \label{fig:method_data}
\end{figure}

\textbf{Storage \& Optimization:} The final processed data were transformed into Analysis-Ready Cloud Optimized (ARCO) Zarr format \citep{newman2024zarr}. The resulting datasets are stored as 3D cubes, chunked into $512 \times 512$ spatial blocks to ensure efficient parallel data retrieval during training (see Figure~\ref{fig:method_data}).

\textbf{Distributional treatment of predictors:} The ISIMIP2b population is heavy-tailed (approximately log-normal across urban--rural gradients), so it is log-transformed ($\ln(x + 10^{-8})$) to prevent the urban tail from dominating training. The aspect variable measures the orientation of the slope of land surface, so it is normalized using a cosine function to preserve its directional nature. All other continuous predictors (LUH2 surface parameters \texttt{secma} and \texttt{secmb}, GEBCO elevation and bathymetry, and PNNL soil composition and slope) are standardised to zero mean and unit variance using statistics computed on the training set. LUH2 fractional land-use variables are already bounded in $[0, 1]$ and require no transformation; categorical inputs (K\"oppen--Geiger climatic zones and the patch-masked autoregressive prior) are entity-embedded and the distributional concern does not apply.

\subsection{Model Architecture}
We frame reconstruction as a per-pixel semantic-segmentation task with eight target labels (the seven land-use classes plus an ocean mask). All models map the stacked predictor tensor---the 14 LUH2 channels, six static geophysical fields, population, and the Köppen-Geiger and (where used) autoregressive priors---to a 1~km land-use map over $256\times256$ patches. Categorical inputs are handled with entity embeddings \citep{guo2016entityembeddingscategoricalvariables}: the 31 Köppen-Geiger classes are embedded into a 5-dimensional vector, and, for autoregressive models, the previous-year HILDA+ map is embedded into a 2-dimensional vector with a dedicated padding index for masked pixels.

We study two segmentation backbones, each in an autoregressive (AR) and a non-\hspace{0pt}autoregressive (non-AR) scheme:

\textit{U-Net.} A convolutional encoder--decoder \citep{ronneberger2015unetconvolutionalnetworksbiomedical} with a 64-channel base that doubles across four downsampling stages (64--128--256--512--512) and mirrors them in the decoder. Its skip connections route high-resolution encoder features directly to the decoder, recovering sharp class boundaries that repeated downsampling would otherwise degrade---a property well suited to resolving kilometre-scale land-use texture from coarse forcings.

\textit{Swin-ViT.} A hierarchical Swin-Transformer encoder (Swin-Tiny: embedding dimension 96, depths $[2,2,6,2]$, $[3,6,12,24]$ attention heads, window size 8) paired with a U-Net-style decoder \citep{cao2021swinunetunetlikepuretransformer}. Self-attention within shifted windows captures long-range spatial dependencies across large land formations that a purely convolutional receptive field may miss.

\textit{Autoregressive prior and training objective.} In the AR scheme, each model is conditioned on the previous year's land-use map and unrolled recurrently, producing sequential predictions for the entire temporal evaluation period. Because land use is strongly temporally autocorrelated, this prior is highly informative; however, feeding the model its own predictions back as input risks compounding error over long rollouts. We mitigate this during training with scheduled sampling, linearly transitioning from teacher forcing toward the model's own predictions. 

The non-AR scheme removes the prior entirely, predicting for each year independently, and thereby trading short-term context for immunity to error accumulation---the central trade-off this paper examines. Year-over-year land-use change is sparse ($\sim$0.53\%/yr globally), so a plain pixel-wise cross-entropy is dominated by the unchanged background and rewards near-persistence solutions. To counter this, all AR models are trained with a cross-entropy objective augmented by a change-weighted term that up-weights pixels whose label differs from the previous year, counteracting the dominance of the static background and focusing capacity on the land-use transitions.

\textit{Baselines.} We compare against a ladder of baselines that isolate successive modelling capabilities: (i) \emph{rolling persistence}, which copies the previous year's map ($\hat{y}_t = y_{t-1}$) and is the trivial no-change reference; (ii) \emph{frozen persistence}, which repeats the last training-year map; (iii) a \emph{linear} model---a per-pixel multinomial logistic regression implemented as a single $1\times1$ convolution ($\sim$430 parameters) with no spatial context, marking the floor for ``no spatial information''; and (iv) a \emph{simple CNN}, an encoder--decoder identical to the U-Net but with its skip connections removed, isolating the contribution of multi-scale skip features. Together these bracket where the deep models' gains originate: temporal modelling (versus persistence), spatial context (versus the linear model), and skip-connected multi-scale fusion (versus the plain CNN).

\begin{figure}[t!]
  \centering
  \includegraphics[width=0.75\textwidth]{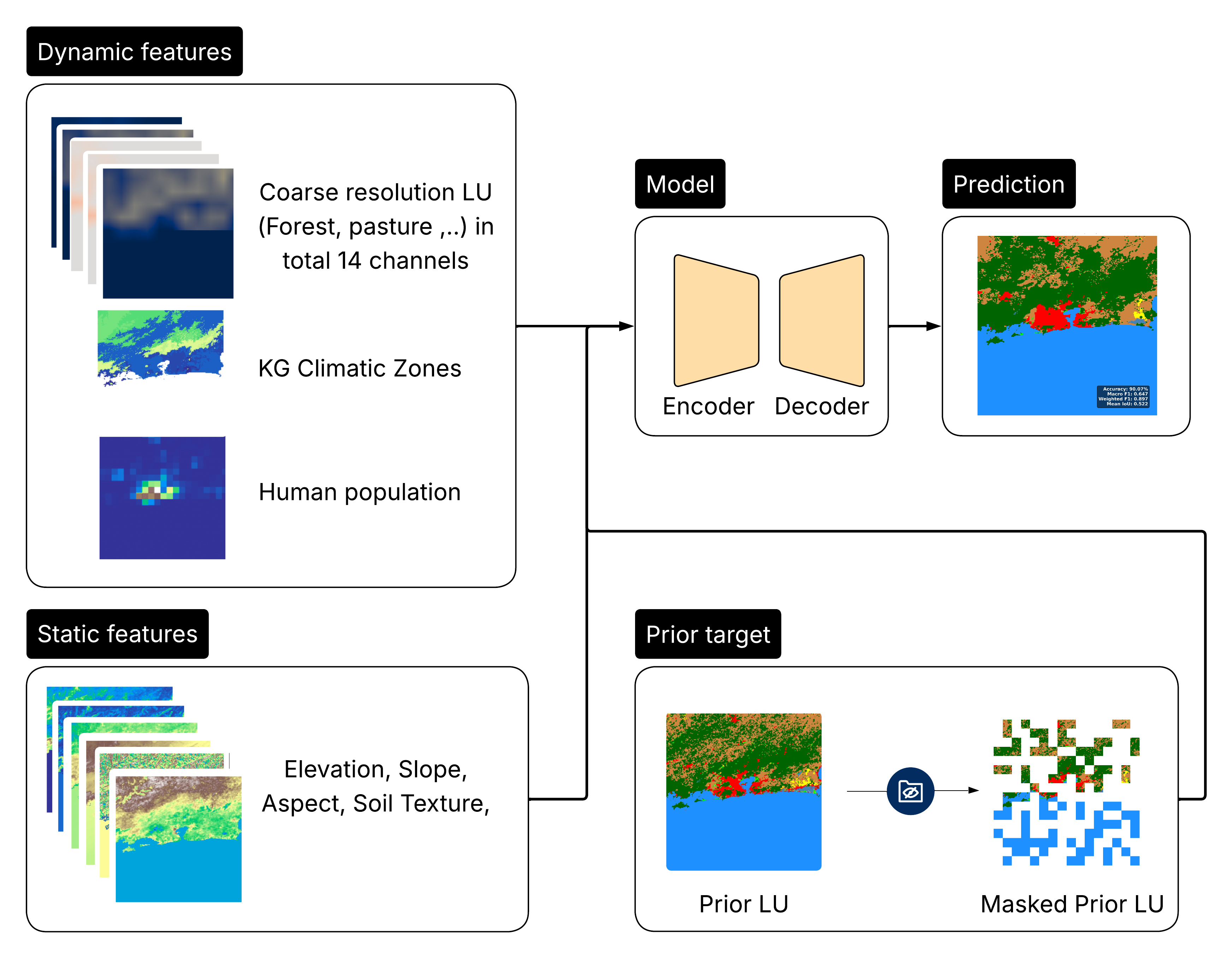}
  \caption{\textbf{Architecture \& Autoregressive Prior.} A schematic of the neural network architecture and the autoregressive training strategy.}
  \label{fig:method_arch}
\end{figure}

\subsection{Evaluation Protocol and Inference}\label{sec:met_protocol}
\textbf{Spatial--temporal splitting.} Land-use maps exhibit strong spatial autocorrelation, so randomly sampling patches would leak neighbouring pixels between training and test and reward regional memorisation rather than genuine generalisation. To prevent this, we partition the globe into coarse macro-grids, each assigned exclusively to the train, validation, or test split; because no macro-grid contributes patches to more than one split, a spatial buffer separates the splits. Within these grids we apply Farthest Point Sampling (FPS) to select 448{,}000 spatially distinct $256 \times 256$ patches (320k train, 64k validation, 64k test). Temporally, models are trained on HILDA+ years 1960--2000 and evaluated on the held-out 2001--2014 period, so the test set is disjoint from training in both space and time \citep{mozaffari2026scalable}. The macro-grid geometry is illustrated in Appendix~\ref{app:split} Each macro-cell spans $9^\circ \times 4.5^\circ$ (approximately $1000 \times 500$~km, i.e.\ ${\sim}5 \times 10^{5}$~km$^2$ at the equator, decreasing towards the poles).

\textbf{Dual-task, change-aware evaluation.} We evaluate two tasks. In \emph{Task~1 (spatial downscaling)} the model predicts year $t$ from that year's coarse drivers alone, measuring pure information transfer from coarse forcing to the 1~km grid. In \emph{Task~2 (autoregressive rollout)} the model is initialised with the ground-truth map of year~2000 and then unrolled in closed loop through 2001--2014, consuming its own previous prediction as the prior, which probes temporal stability and error accumulation. Because the vast majority of pixels do not change between consecutive years, aggregate scores are dominated by the static background; we therefore report every metric in two regimes---\emph{overall} (all pixels) and \emph{land-use change} (pixels whose label differs from the previous year)---the latter being the scientifically meaningful signal on which trivial rolling persistence scores zero by construction. We report pixel accuracy, macro- and weighted-F1, and mean Intersection-over-Union (mIoU), together with per-class scores, averaged over five random seeds.

\textbf{Inference.} To generate seamless global maps, we employ a sliding window approach with a Gaussian weighting function $W$ to merge overlapping predictions \citep{mozaffari2026scalable}. The final probability $\bar{\mathbf{P}}(x, y)$ is computed by normalizing the weighted sum of overlapping patches $\Omega_{(x,y)}$:
\begin{equation}
\bar{\mathbf{P}}(x, y) = \frac{\sum_{k \in \Omega} \mathbf{p}^{(k)}(x, y) \cdot W(u_k, v_k)}{N(x, y)}
\end{equation}
The final class is derived via $\operatorname*{argmax}_{c}$, ensuring smooth transitions across patch boundaries without ``blocky`` artifacts.

\subsection{The TerraNostra-Bench Benchmark}
We package the data, splits, tasks, metrics, and baselines defined above into \textbf{TerraNostra-Bench}, a reproducible benchmark for high-resolution land-use reconstruction. It standardises: (i) the \emph{target}, HILDA+ consolidated to seven land-use classes plus an ocean mask at 1~km; (ii) the \emph{predictors}, the harmonised LUH2, Köppen-Geiger, population, and static geophysical fields on a common grid; (iii) the \emph{spatial--temporal split}, the fixed macro-grid partition (320k/64k/64k patches) with training on 1960--2000 and testing on the held-out 2001--2014 period; (iv) the \emph{two evaluation tasks}, spatial downscaling (Task~1) and the 14-year autoregressive rollout (Task~2), each scored in the overall and land-use-change regimes; and (v) a set of \emph{reference baselines}---rolling and frozen persistence, a linear model, and a skip-free CNN---against which any submission is compared. By fixing these choices, TerraNostra-Bench enables direct, leakage-free comparison of land-use downscaling models, which existing remote-sensing benchmarks---designed for static classification or short-horizon regional segmentation---do not support. The dataset and evaluation code will be released publicly upon publication at \url{https://zenodo.org/records/23040508}.

\section{Results}

We evaluate the four learned variants---U-Net and Swin-ViT, each in autoregressive (AR) and non-\hspace{0pt}autoregressive (non-AR) scheme---together with the persistence, linear, and convolutional baselines, on Tasks~1 and~2. Unless stated otherwise, all metrics are means over five random seeds and averaged across the 2001--2014 test horizon; per-seed standard deviations are below $0.01$ for every aggregate metric.

\begin{figure}[ht!]
  \centering
  \includegraphics[width=0.62\linewidth]{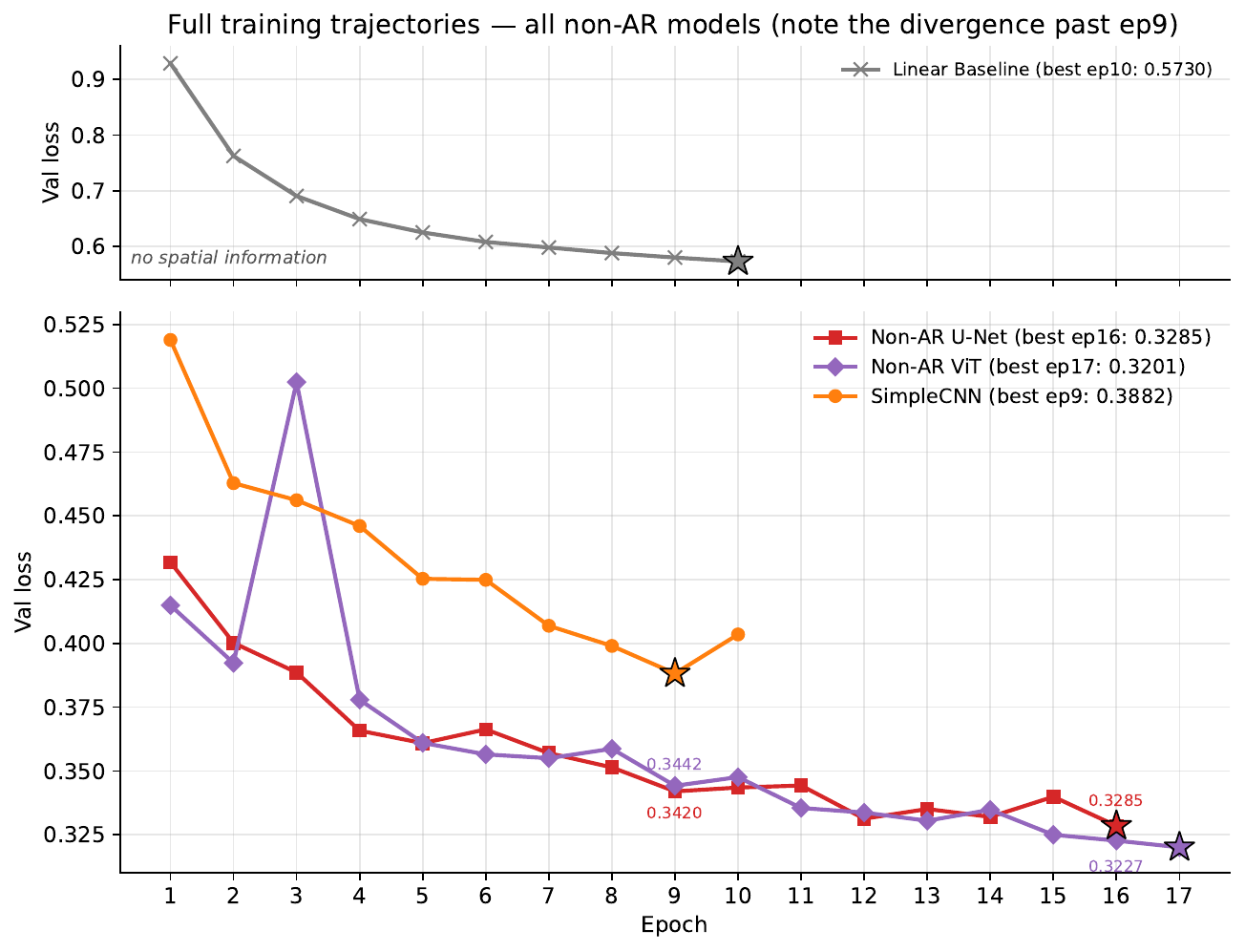}
  \caption{\textbf{Validation-loss convergence} for the non-\hspace{0pt}autoregressive runs and the learnable baselines. The linear model (top panel) at a much higher validation loss ($\approx 0.57$) than the deep models ($0.32$--$0.39$, bottom panel), quantifying the value of spatial context.}
  \label{fig:trainloss}
\end{figure}

Figure~\ref{fig:trainloss} summarises validation-loss convergence for the non-\hspace{0pt}autoregressive training runs and the learnable baselines: the deep models plateau between $0.32$ and $0.39$, well below the linear model's plateau ($0.57$), outperforming the baseline by a long shot and indicating that spatial context is essential. To enable comparison at a matched training budget we mark epoch~9, the shared horizon of the ViT and CNN runs, and report the non-\hspace{0pt}autoregressive U-Net at this epoch-9 checkpoint rather than its later optimum, for a like-for-like comparison. The linear and simple-CNN baselines complete the architecture ladder: both are now evaluated over the full five-seed protocol and are reported alongside the deep models throughout.

\subsection{Overall versus change-aware performance}\label{sec:res_overall}

Table~\ref{tab:overall_change} reports each model in the two regimes introduced in Section~\ref{sec:met_protocol}: \emph{overall} (all pixels) and \emph{land-use change} (only pixels whose label differs from the previous year). The two regimes tell opposite stories, which resolves an apparent paradox: judged on overall accuracy, the trivial \emph{rolling persistence} baseline appears best at $0.9947$, simply because $\sim$98\% of pixels are unchanged year-to-year and copying the previous map scores them all correctly. Rolling persistence is, moreover, not a usable reconstruction method: it requires the observed map of the previous year, so it can only produce a single year beyond the last observation, not the multi-decadal projections to 2100 that motivate this work. On the land-use-change pixels that actually matter, however, rolling persistence scores exactly zero by construction, and frozen persistence only $0.16$, whereas the learned models reach $0.35$--$0.39$. This frozen-persistence score does not reflect forecasting skill but a structural property of land-use dynamics: some changes are reversions, for example when a temporarily deforested pixel regrows to forest, or when a fallow-cropland cycle returns to the year-2000 land use. The $0.16$ therefore reflects the fraction of LU change events that are reversions to the year-2000 state, a property of the data, not a sign that the baseline is forecasting anything.

\begin{table}[ht!]
    \centering
    \caption{\textbf{Overall versus land-use-change performance} (5-seed means, 2001--2014 average). Overall accuracy is dominated by the pixels that stay in the same class between consecutive years, so persistence baselines are unbeatable on this axis; the land-use-change regime---the scientifically meaningful signal---measures accuracy only on the pixels that actually transitioned. The two regimes separate the models cleanly: the non-\hspace{0pt}autoregressive variants lead on overall accuracy and mIoU, the autoregressive variants---trained with a change-weighted objective---lead on every change metric. The Linear and Simple CNN baselines obtain generally worse results across the metrics. mIoU is the macro-average of per-class IoU over the 7 land classes (Ocean excluded, as it is trivially separable at IoU $\approx$ 0.999 for every model). Accuracy columns are pixel-level micro averages. The two non-\hspace{0pt}autoregressive overall-mIoU scores (0.5276, 0.5264) are within seed noise of each other.}
    \label{tab:overall_change}
    \small
    \begin{tabular}{lcccc}
        \toprule
        & \multicolumn{2}{c}{\textbf{Overall}} & \multicolumn{2}{c}{\textbf{Land-use change}} \\
        \cmidrule(lr){2-3}\cmidrule(lr){4-5}
        \textbf{Model} & \textbf{Acc.} & \textbf{mIoU} & \textbf{Acc.} & \textbf{mIoU} \\
        \midrule
        Rolling persistence   & 0.9947 & 0.9890 & 0.0000 & 0.0000 \\
        Frozen persistence    & 0.9704 & 0.9419 & 0.1630 & 0.0883 \\
        Linear                & 0.8058 & 0.3988 & 0.3510 & 0.1458 \\
        Simple CNN            & 0.8193 & 0.4879 & 0.3603 & 0.1466 \\
        \midrule
        AR U-Net              & 0.8072 & 0.4218 & \textbf{0.3931} & 0.1661 \\
        non-AR U-Net          & 0.8399 & \textbf{0.5276} & 0.3506 & 0.1425 \\
        AR Swin-ViT           & 0.7923 & 0.4417 & 0.3857 & \textbf{0.1728} \\
        non-AR Swin-ViT       & \textbf{0.8455} & 0.5264 & 0.3648 & 0.1551 \\
        \bottomrule
    \end{tabular}
\end{table}

A second pattern is visible within the learned models: the AR variants achieve the highest land-use-change accuracy $(0.39)$, while the non-AR variants achieve the highest overall accuracy and mIoU. This contrast reflects a deliberate difference in training objective
rather than an intrinsic property of the scheme. The AR models are trained with the
change-weighted objective ($\alpha=0.5$), spending half the loss budget on the rare
transition pixels; while the non-AR models are not passed a prior to compare against the target, so the training is based on pure reconstruction ($\alpha=1.0$).The AR models thus trade overall
accuracy for change sensitivity by construction. Trained on the \emph{same}
pure-reconstruction objective, they in fact surpass every non-AR variant on overall accuracy
(AR U-Net $0.8971$, AR Swin-ViT $0.8594$ at $\alpha=1.0$; Section~\ref{sec:lossbalance}),
confirming that the autoregressive prior is the more informative input and that its apparent
overall deficit here is a chosen operating point, not a limitation. What the AR scheme genuinely costs is long-horizon stability---the trade-off we dissect next.

\subsection{Spatial consistency}\label{sec:res_spatial}
To assess the spatial consistency and robustness of the spatial–temporal splitting protocol (Section~\ref{sec:met_protocol}), we evaluated whether boundary artifacts affect inference on $256 \times 256$ patches.

Since macro-grid sampling enforces hard spatial boundaries to eliminate data leakage, truncating peripheral context could theoretically degrade performance near patch edges. We therefore performed an inference-time ablation using symmetric spatial padding of 16 and 32 pixels across the full 14-year rollout (2001–2014).

As shown in Table~\ref{tab:spatial_consistency}, restoring perimeter context has a negligible impact across all architectures: variations in accuracy and mIoU remain bounded between $-0.53\%$ and $+0.31\%$ for both $\Delta(\mathrm{Padding}_{16} - \mathrm{Padding}_{0})$ and $\Delta(\mathrm{Padding}_{32} - \mathrm{Padding}_{0})$.

These results confirm that the protocol using $256 \times 256$ patches maintains the spatial consistency required for land-use projections, without being compromised by the spatial isolation used to counteract autocorrelation.

\begin{table}[ht!]
    \centering
    \caption{\textbf{Padding effect on inference overall 14-Year rollout mean} (3-seed means, 2001-2014 average).}
    \label{tab:spatial_consistency}
    \small
    \begin{tabular}{lcccc}
        \toprule
        & \multicolumn{2}{c}{\textbf{$\Delta(\mathrm{Padding}_{16} - \mathrm{Padding}_{0})$}} & \multicolumn{2}{c}{\textbf{$\Delta(\mathrm{Padding}_{32} - \mathrm{Padding}_{0})$}} \\
        \cmidrule(lr){2-3}\cmidrule(lr){4-5}
        \textbf{Model} & \textbf{Acc.} & \textbf{mIoU} & \textbf{Acc.} & \textbf{mIoU} \\
        \midrule
        Linear                & -0.17\% & +0.15\% & -0.16\% & -0.18\% \\
        Simple CNN            & +0.02\% & +0.20\% & -0.13\% & -0.24\% \\
        \midrule
        AR U-Net              & -0.13\% & \textbf{-0.53}\% & -0.19\% & -0.38\% \\
        non-AR U-Net          & -0.01\% & -0.39\% & -0.00\% & -0.18\% \\
        AR Swin-ViT           & +0.17\% & +0.20\% & \textbf{+0.31}\% & +0.08\% \\
        non-AR Swin-ViT       & -0.10\% & +0.08\% & -0.13\% & +0.18\% \\
        \bottomrule
    \end{tabular}
\end{table}

\subsection{Temporal stability and the autoregressive trade-off}\label{sec:res_rollout}
\begin{figure}[ht!]
  \centering
  \includegraphics[width=0.7\linewidth]{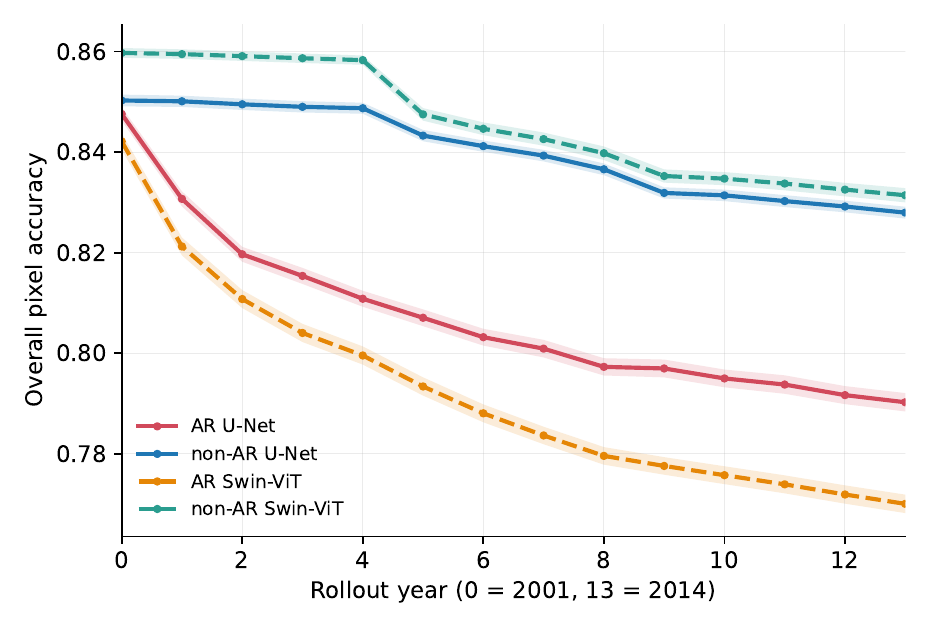}
  \caption{\textbf{Overall accuracy across the 14-year autoregressive rollout} (5-seed mean, shaded $\pm$1\,s.d.). All models decay, but autoregressive models show a much steeper decline than non-\hspace{0pt}autoregressive models.}
  \label{fig:rollout_acc}
\end{figure}

Figure~\ref{fig:rollout_acc} traces overall accuracy across the 14-year closed-loop rollout (Task~2), and Table~\ref{tab:rollout} summarises its endpoints. We can see how the accuracy of the autoregressive models decays as self-conditioning compunds error: AR U-Net loses $5.7$ points and AR Swin-ViT $7.2$ points by year~13. The non-AR models, which predict each year independently, have much more stable results (non-AR U-Net $-2.2$ points, non-AR Swin-ViT $-2.8$ points). Within each scheme, the Swin-ViT degrades faster than the U-Net: the non-\hspace{0pt}autoregressive Swin-ViT starts highest, but its lead over the U-Net shrinks from $0.9$ to $0.3$ points by year~13, and in the autoregressive setting the U-Net leads throughout. This suggests that the convolutional inductive bias is more robust over long horizons, whereas the Swin-ViT's attention helps mainly at short range; the backbone effect nonetheless remains secondary to the autoregressive choice. For autoregressive models trained with the change-weighted objective, a recent ground-truth prior therefore brings little overall benefit and becomes detrimental over long horizons, due to the accumulated error. This cautions us against naive autoregressive rollout for the multi-decadal projections this framework ultimately targets.

\begin{table}[ht!]
    \centering
    \caption{\textbf{Rollout endpoints} (5-seed means $\pm$ s.d.). Overall accuracy at the first (2001) and last (2014) rollout year, and the change across the horizon. All models decline, including the Linear baseline, which carries no temporal state at all and therefore isolates the component of the decline attributable to the test years themselves rather than to error accumulation ($-0.010$). The autoregressive models lose a further $4$--$6$ points beyond that floor, whereas the non-\hspace{0pt}autoregressive models lose only $1$--$2$. Bold marks the best value among the four deep models.}
    \label{tab:rollout}
    \small
    \begin{tabular}{lccc}
        \toprule
        \textbf{Model} & \textbf{Year 0} & \textbf{Year 13} & \textbf{$\Delta$ (Y0$\to$Y13)} \\
        \midrule
         AR U-Net        & $0.8476 \pm 0.0013$ & $0.7903 \pm 0.0018$ & $-0.057$ \\
        AR Swin-ViT     & $0.8421 \pm 0.0016$ & $0.7701 \pm 0.0018$ & $-0.072$ \\
        \midrule
        non-AR U-Net    & $0.8503 \pm 0.0012$ & $0.8280 \pm 0.0012$ & $\mathbf{-0.022}$ \\
        non-AR Swin-ViT & $\mathbf{0.8597} \pm 0.0010$ & $\mathbf{0.8314} \pm 0.0014$ & $-0.028$ \\
        \midrule
        Simple CNN      & $0.8323 \pm 0.0010$ & $0.8073 \pm 0.0012$ & $-0.025$ \\
        Linear          & $0.8105 \pm 0.0012$ & $0.8004 \pm 0.0012$ & $-0.010$ \\
        \bottomrule
    \end{tabular}
\end{table}

\begin{figure}[ht!]
  \centering
  \includegraphics[width=0.7\linewidth]{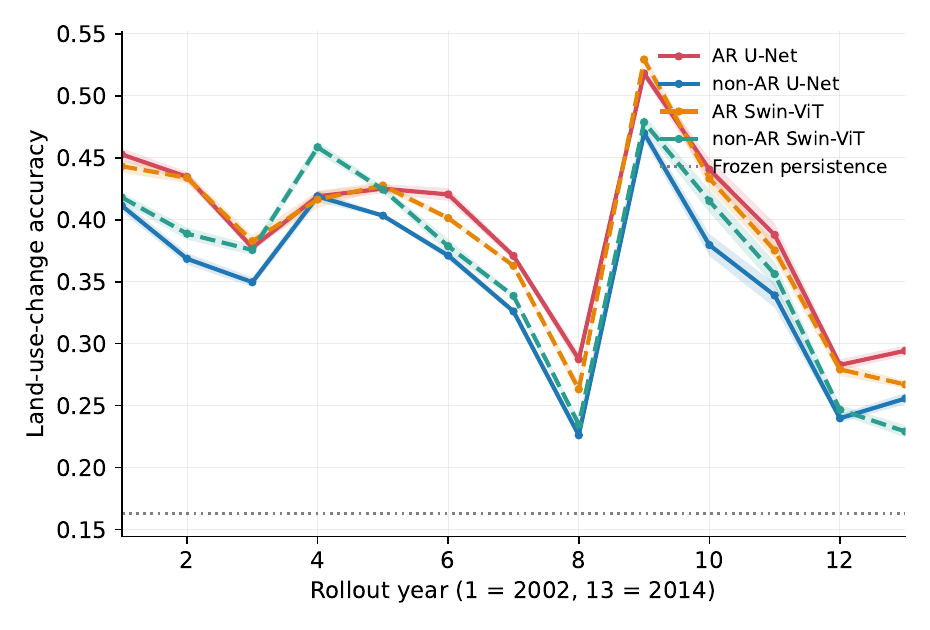}
  \caption{\textbf{Land-use-change accuracy across the rollout} (5-seed mean, shaded $\pm$1\,s.d.). All learned models far exceed the frozen-persistence floor (dotted); autoregressive variants lead on change detection.}
  \label{fig:change_acc}
\end{figure}

The land-use-change view of the same rollout (Figure~\ref{fig:change_acc}) shows that every learned model stays far above the frozen-persistence floor ($0.16$). Detecting where the map changes is precisely what persistence cannot do, and it is where the learned models earn their keep. In general, we see that AR models obtain higher change-detection accuracy across the horizon, as opposed to their non-\hspace{0pt}autoregressive counterparts.

Qualitative rollout examples for representative high- and moderate-change patches are provided in Appendix~\ref{app:drift}.

\subsection{Feature importance}\label{sec:res_ablation}
To quantify the contribution of each input, we performed a leave-one-out study, removing one feature (or feature group) at a time and measuring the change in validation loss relative to the full model; a larger increase indicates a more important input. Table~\ref{tab:ablation} reports the result for the autoregressive U-Net; we then repeat the study without the prior (Table~\ref{tab:ablation_non_ar}) to separate what the drivers contribute from what the prior supplies. The autoregressive HILDA+ prior dominates by a wide margin: removing it raises the loss by $16.7\%$, more than twice the effect of any other input, confirming that the previous-year map is the single most informative signal---and motivating the AR-versus-non-AR study above. Among the remaining predictors, the non-forested LUH2 states (\texttt{secdn}, $+7.7\%$ and \texttt{primn}, $+5.8\%$), C4 annual crops (\texttt{c4ann}, $+5.3\%$), the dynamic Köppen-Geiger class ($+4.8\%$), and population ($+3.3\%$) contribute most. Removing the static-feature or full-LUH2 groups leaves the  validation loss essentially unchanged, a result whose cause becomes clear once the prior is
removed (Table~\ref{tab:ablation_non_ar}).

\begin{table}[ht!]
    \centering
    \caption{\textbf{Leave-one-out feature importance.} Change in validation loss when each input (or group) is removed, relative to the full model (loss $0.5871$); a positive $\Delta$ means the feature is helpful. Köppen-Geiger and population are supplied as dynamic annual inputs. The autoregressive prior dominates.}
    \label{tab:ablation}
    \small
    \begin{tabular}{lccc}
        \toprule
        \textbf{Removed input} & \textbf{Val. loss} & \textbf{$\Delta$} & \textbf{\% change} \\
        \midrule
        \textbf{Autoregressive HILDA+ prior} & \textbf{0.6854} & \textbf{$+0.0983$} & \textbf{$+16.74$} \\
        Secondary non-forested (\texttt{secdn})    & 0.6322 & $+0.0451$ & $+7.68$ \\
        Primary non-forested (\texttt{primn})      & 0.6210 & $+0.0339$ & $+5.77$ \\
        C4 annual crops (\texttt{c4ann})           & 0.6181 & $+0.0310$ & $+5.28$ \\
        Köppen-Geiger class (dynamic)              & 0.6151 & $+0.0280$ & $+4.77$ \\
        Secondary deciduous forest (\texttt{secdf})& 0.6096 & $+0.0225$ & $+3.83$ \\
        Population (dynamic)                       & 0.6065 & $+0.0194$ & $+3.30$ \\
        \midrule
        All LUH2 features                          & 0.5840 & $-0.0031$ & $-0.53$ \\
        All static features                        & 0.5707 & $-0.0164$ & $-2.79$ \\
        \bottomrule
    \end{tabular}
\end{table}

The same study repeated on the non-\hspace{0pt}autoregressive U-Net (Table~\ref{tab:ablation_non_ar})
reveals how much the prior was masking. With no previous-year map to lean on, the dynamic
Köppen--Geiger class becomes the single indispensable input: removing it nearly doubles the validation loss ($+95.7\%$), against $+4.8\%$ in the autoregressive setting. The group-level
pattern inverts just as sharply. In the AR model, deleting \emph{all} LUH2 or \emph{all} static features left the loss essentially unchanged (indeed marginally lower); in the non-AR
model the same deletions cost $+26.8\%$ and $+29.8\%$ respectively. Individual LUH2 channels, by contrast, remain nearly free to remove ($-2.1\%$ to $+1.9\%$), as expected given that the fractional state variables are compositional and largely recoverable from one another---the information is carried by the group, not by any single channel. Because the two models are trained with different objectives, we compare relative degradations rather than absolute
losses.

\begin{table}[ht!]
    \centering
    \caption{\textbf{Leave-one-out feature importance (non-\hspace{0pt}autoregressive U-Net).} Change in
    validation loss when each input (or group) is removed, relative to the full non-AR model
    (loss $0.4210$); a positive $\Delta$ means the feature is helpful. Without the autoregressive HILDA+ prior,
    the dynamic Köppen-Geiger class dominates, and removing whole feature groups is far more
    damaging than removing any single LUH2 channel. Absolute losses are not comparable to
    Table~\ref{tab:ablation}, as the non-AR model is trained without the change-loss term.}
    \label{tab:ablation_non_ar}
    \small
    \begin{tabular}{lccc}
        \toprule
        \textbf{Removed input} & \textbf{Val. loss} & \textbf{$\Delta$} & \textbf{\% change} \\
        \midrule
        \textbf{Köppen-Geiger class (dynamic)} & \textbf{0.8237} & \textbf{$+0.4027$} & \textbf{$+95.65$} \\
        Primary non-forested (\texttt{primn})       & 0.4291 & $+0.0081$ & $+1.92$ \\
        Secondary deciduous forest (\texttt{secdf}) & 0.4232 & $+0.0022$ & $+0.52$ \\
        Population (dynamic)                        & 0.4211 & $+0.0001$ & $+0.02$ \\
        Secondary non-forested (\texttt{secdn})     & 0.4131 & $-0.0079$ & $-1.88$ \\
        C4 annual crops (\texttt{c4ann})            & 0.4123 & $-0.0087$ & $-2.07$ \\
        \midrule
        All static features                         & 0.5464 & $+0.1254$ & $+29.79$ \\
        All LUH2 features                           & 0.5337 & $+0.1127$ & $+26.77$ \\
        \bottomrule
    \end{tabular}
\end{table}

Together the two ablations give a mechanistic account of the trade-off documented above. The autoregressive prior is not merely the most informative input; it is \emph{substitutive},
supplying enough of the signal that the physical drivers become individually and collectively redundant. Strip it away and the model must infer land use from climate, terrain, and coarse land-use forcing, at which point those drivers become essential. This is precisely why the AR variants degrade under rollout: they have learned to rely on a signal that, once self-generated, carries compounding error, whereas the non-AR variants stay anchored to drivers that remain exogenous and error-free at every step.

\subsection{Per-class performance}\label{sec:res_class}
\begin{figure}[ht!]
  \centering
  \includegraphics[width=0.85\linewidth]{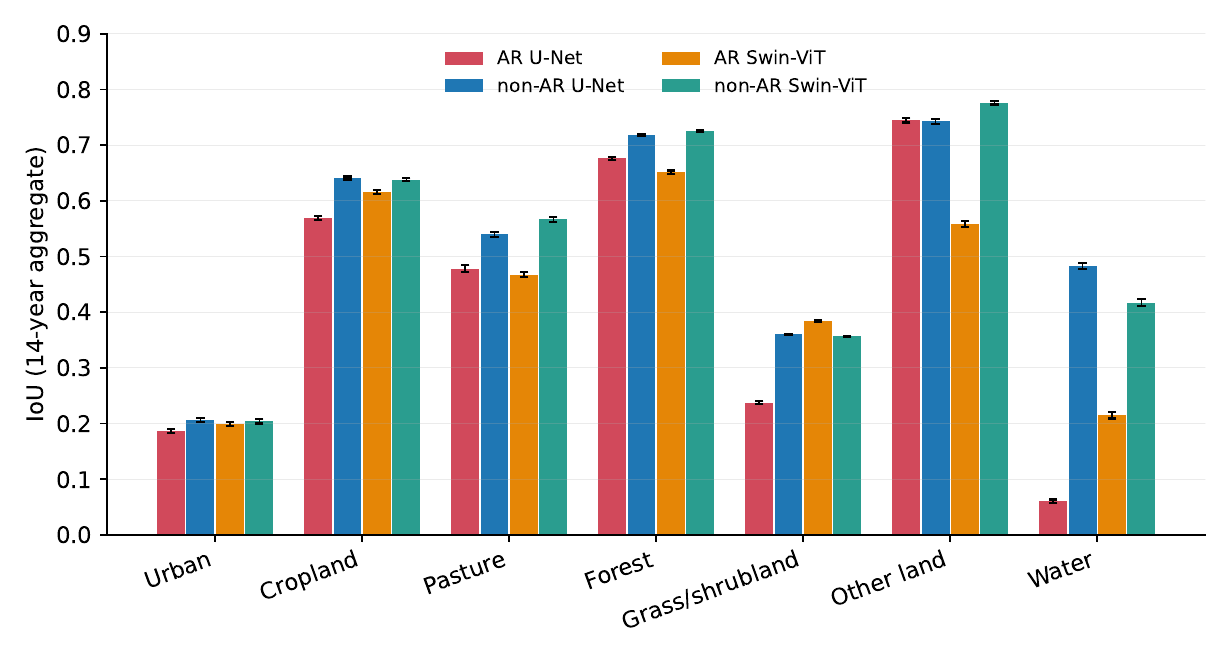}
  \caption{\textbf{Per-class IoU} across the four learned variants (5-seed mean, error bars $\pm$1\,s.d., 2001--2014 aggregate). The autoregressive Water collapse and the difficulty of the rare Urban and Grass/Shrubland classes are visible, while the non-\hspace{0pt}autoregressive variants lead on most classes.}
  \label{fig:perclass_bar}
\end{figure}

Table~\ref{tab:perclass} and Figure~\ref{fig:perclass_bar} break the 14-year-average reconstruction down by class (IoU). The non-\hspace{0pt}autoregressive variants achieve the highest IoU on most classes, a direct consequence of their flat rollout: because the AR models degrade year-on-year, their class-averaged scores fall behind. The gap is starkest for Water---the inland-water class of lakes, rivers and reservoirs---where AR U-Net collapses to $0.06$ while non-AR U-Net retains $0.48$. This class is the one for which the coarse forcing offers no direct evidence: none of the 14 LUH2 channels encodes open water, so water can only be inferred from the static geophysical fields or carried over from the previous year's map. The autoregressive variants rely on the latter and therefore lose the class as the prior degrades over the rollout, whereas the non-\hspace{0pt}autoregressive variants, grounded in the static fields alone, retain it. Rare, spatially fragmented classes remain hardest for all models: Urban (IoU $0.19$--$0.21$) and Grass/Shrubland ($0.24$--$0.38$) are limited by their small footprint and diffuse boundaries, whereas spatially coherent classes such as Forest are reconstructed well ($>0.68$). Within each scheme the two backbones are otherwise comparable, indicating that the autoregressive choice, not the U-Net-versus-Transformer backbone, is the primary driver of per-class quality.

\begin{table}[ht!]
    \centering
    \caption{\textbf{Per-class IoU} (5-seed mean, 2001--2014 average) for the four learned variants. Non-\hspace{0pt}autoregressive models lead on six of the seven classes; the AR Water collapse reflects rollout error accumulation on the one class for which the coarse forcing carries no direct evidence. The ocean class is excluded.}
    \label{tab:perclass}
    \small
    \begin{tabular}{lcccc}
        \toprule
        \textbf{Class} & \textbf{AR U-Net} & \textbf{non-AR U-Net} & \textbf{AR Swin-ViT} & \textbf{non-AR Swin-ViT} \\
        \midrule
        Urban           & 0.1864 & \textbf{0.2068} & 0.1994 & 0.2043 \\
        Cropland        & 0.5687 & \textbf{0.6410} & 0.6159 & 0.6381 \\
        Pasture         & 0.4785 & 0.5401 & 0.4676 & \textbf{0.5668} \\
        Forest          & 0.6760 & 0.7186 & 0.6512 & \textbf{0.7255} \\
        Grass/shrubland & 0.2376 & 0.3605 & \textbf{0.3843} & 0.3568 \\
        Other land      & 0.7447 & 0.7430 & 0.5581 & \textbf{0.7763} \\
        Water           & 0.0609 & \textbf{0.4829} & 0.2154 & 0.4171 \\
        \bottomrule
    \end{tabular}
\end{table}

\section{Controlling the autoregressive trade-off}\label{sec:control}

The autoregressive trade-off dissected above is not a fixed property of a trained model: it can be moved along the stability--sensitivity frontier by two complementary levers. The first acts at \emph{inference-time}, requiring no retraining, a per-pixel trust threshold that decides, at each rollout step whether to accept the model's own prediction or fall back to the carried-over prior. The second acts at \emph{training-time}, the relative weighting of the overall-reconstruction and land-use-change loss terms. We characterise both, show that they trace overlapping but non-identical frontiers, and give practical guidance on when to reach for each.

\subsection{Inference-time control: prior-trust calibration}\label{sec:trust}

\begin{figure}[ht!]
  \centering
  \includegraphics[width=0.72\linewidth]{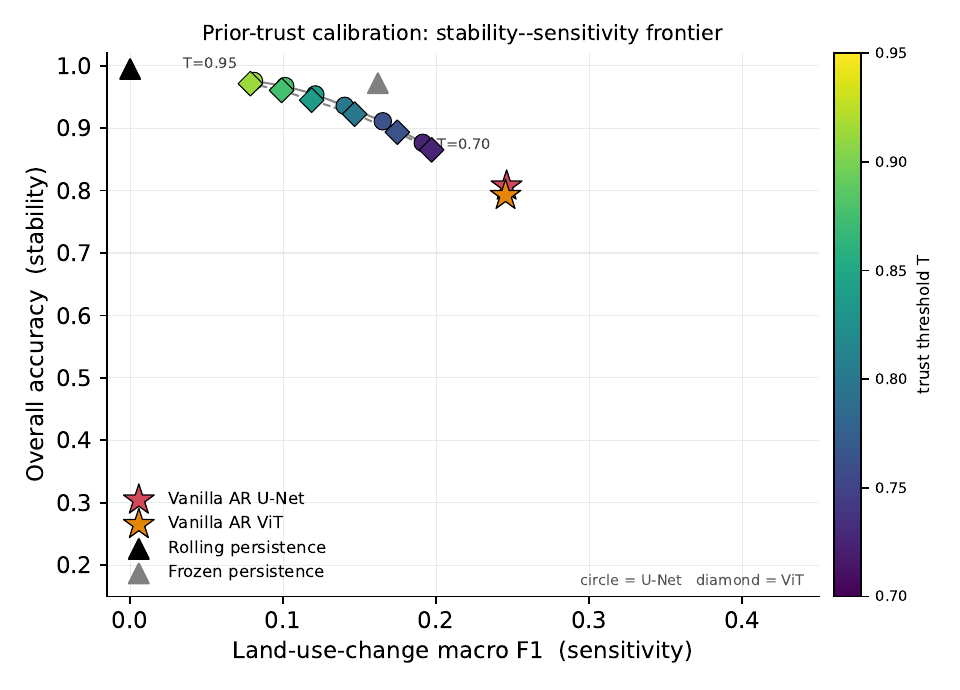}
  \caption{\textbf{Prior-trust calibration traces a stability--sensitivity frontier.}
  Overall accuracy (stability) versus land-use-change macro-F1 (sensitivity) over the
  14-year rollout, as the inference-time trust threshold is swept from 0.70 to 0.95
  (circles: AR U-Net; diamonds: AR Swin-ViT; colour: threshold). Stars mark the vanilla
  autoregressive models (5-seed mean) and triangles the persistence baselines. Raising the
  threshold moves each model from the change-sensitive vanilla corner toward the stable
  persistence corner; the two architectures trace nearly the same curve. 
  }
  \label{fig:trust_pareto}
\end{figure}

At every rollout step the autoregressive model emits a per-pixel class distribution. We define a pixel's \emph{change probability} as one minus the probability it assigns to the carried-over prior class, and accept the model's prediction only where this exceeds a trust threshold $T$; elsewhere, the pixel retains its prior class. The threshold is a single scalar applied at inference to an already-trained checkpoint---no retraining, and the same mechanism for both backbones. Sweeping $T$ from $0.70$ to $0.95$ traces the frontier in Figure~\ref{fig:trust_pareto}, moving each model from the change-sensitive vanilla corner toward the stable persistence corner.

As the threshold $T$ increases, the effect becomes more similar across architectures. At $T=0.75$, overall 14-year rollout accuracy rises from $0.8072$ to $0.9110$ for the AR U-Net ($+10.4$ points) and from $0.7923$ to $0.8935$ for the AR Swin-ViT ($+10.1$ points), while the year-0-to-year-13 accuracy drift is cut by $28\%$ (U-Net) and $19\%$ (Swin-ViT).  Despite this gain in stability, land-use-change macro-F1 also remains above the frozen-persistence floor ($0.162$). The two architectures trace closely parallel curves that converge to within $0.01$ on overall accuracy and land-use-change macro-F1 at high thresholds ($T\!\ge\!0.85$), with a still small divergence (up to $0.02$ on accuracy metrics) at the lower thresholds. The calibration is therefore best read as architecture-agnostic in shape, letting a practitioner dial in a stability--sensitivity operating point post hoc without touching the weights.

\subsection{Training-time control: reconstruction--change loss balancing}\label{sec:lossbalance}

\begin{figure}[ht!]
  \centering
  \includegraphics[width=0.72\linewidth]{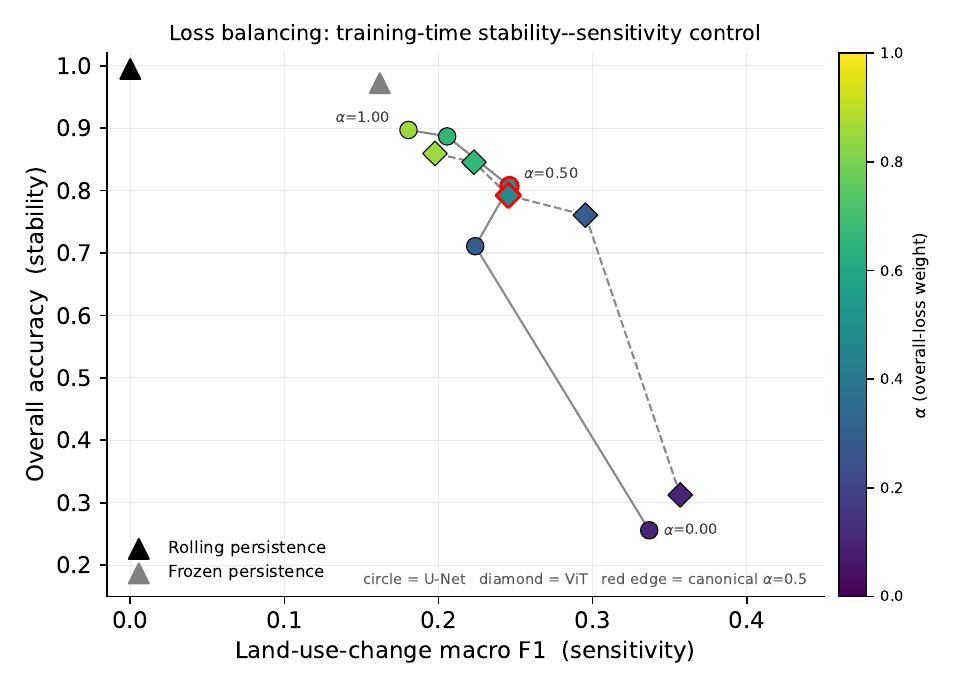}
  \caption{\textbf{Training-time loss balancing traces a stability--sensitivity frontier.} Overall 14-year rollout accuracy (stability) versus land-use-change macro-F1 (sensitivity) as the loss weight $\alpha$ is swept from $0$ to $1$ (circles: U-Net; diamonds: Swin-ViT; colour: $\alpha$), five seeds per point. Increasing $\alpha$ shifts weight onto the overall-reconstruction term, moving each model from the change-sensitive corner toward the stable-reconstruction corner; the canonical default $\alpha=0.5$ (red edge) sits near the knee. Triangles mark the persistence baselines.}
  \label{fig:loss_pareto}
\end{figure}

The training objective combines an overall-reconstruction cross-entropy and a land-use-change cross-entropy, weighted $\alpha$ and $1-\alpha$; the models above use the balanced default $\alpha=0.5$. We retrained both backbones from scratch across $\alpha\in\{0,0.25,0.5,0.75,1.0\}$ (five seeds each) to map how this weighting moves the frontier (Figure~\ref{fig:loss_pareto}). Overall rollout accuracy rises monotonically with $\alpha$: at $\alpha=1.0$ (pure reconstruction) it reaches $0.8971$ for the U-Net ($+9.0$ points over the balanced default) and $0.8594$ for the Swin-ViT ($+6.7$ points), though still well short of the persistence ceiling---confirming the prior, not the change-loss term, carry most of the reconstruction signal. Change-detection, by contrast, is \emph{not} monotonic in $\alpha$ and its optimum is metric- and architecture-dependent; the degenerate $\alpha=0$ extreme barely learns the base distribution (overall accuracy $0.26$--$0.31$) for no reliable change-detection gain. The balanced $\alpha=0.5$ sits near the knee of the frontier for both backbones---no single weighting dominates it on both regimes simultaneously---which validates it as a sensible default.

\subsection{Complementarity of the two levers}\label{sec:combined_control}

\begin{figure}[ht!]
  \centering
  \includegraphics[width=0.72\linewidth]{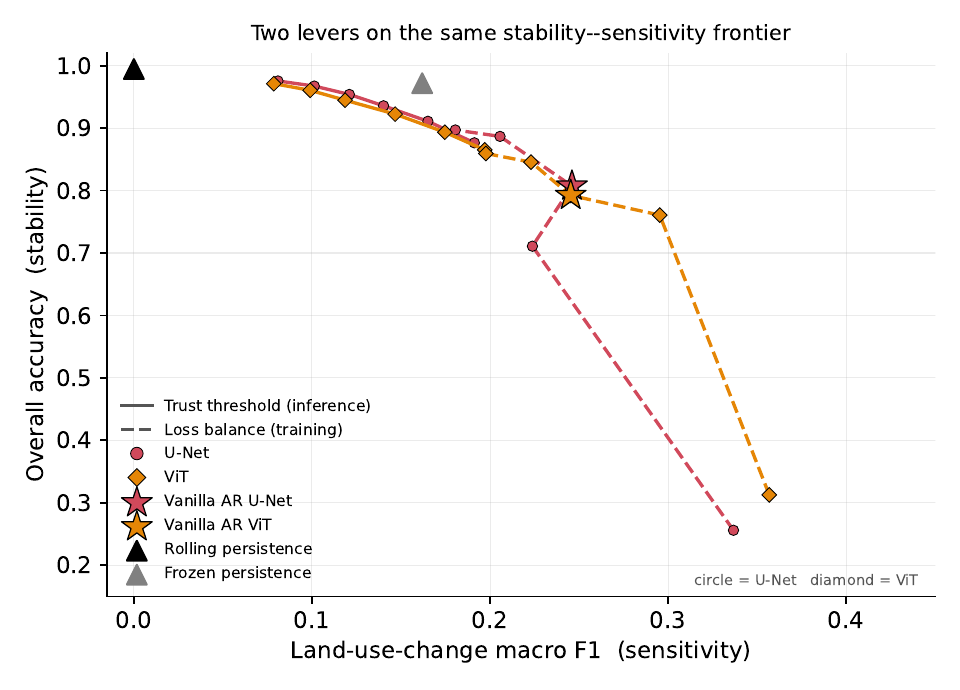}
  \caption{\textbf{The two control levers trace overlapping but non-identical frontiers.} Overall rollout accuracy (stability) versus land-use-change macro-F1 (sensitivity), overlaying the inference-time trust-threshold sweep (solid) and the training-time loss-balance sweep (dashed); circles denote U-Net and diamonds Swin-ViT. Stars mark the vanilla autoregressive models ($\alpha=0.5$, $T=0$) and triangles the persistence baselines. The two levers coincide in the middle of the frontier but each reaches a region the other cannot: the trust threshold extends further toward the stable persistence corner, loss-balancing further toward the change-sensitive corner.}
  \label{fig:unified_pareto}
\end{figure}

Overlaying the two levers (Figure~\ref{fig:unified_pareto}) shows they are \emph{complementary}, not redundant. Note that the levers are compared as alternatives rather than merged into a single model: each loss-balancing point is a separately trained model, whereas the trust threshold is applied post hoc to one fixed checkpoint. They coincide in the middle of the frontier ($T\!\in\![0.70,0.75]$ and $\alpha\!\in\![0.75,1.0]$ land on the same curve), but each reaches a region the other cannot: the inference-time trust threshold can push arbitrarily far toward the stable persistence corner (it can always fall back to the prior), whereas loss-balancing can push further toward maximum change-sensitivity than any threshold (only a retrained, change-weighted model exceeds the vanilla checkpoint's own change ceiling). The practical implication is a decision rule: for maximum stability with no retraining, raise the trust threshold; for maximum change-sensitivity, down-weight $\alpha$ and retrain; for a balanced product, the uncalibrated $\alpha=0.5$ model already sits at the knee of both frontiers. Combining both levers in a single model is left for future work.

\section{Out-of-window extrapolation}\label{sec:res_extrap}

We now probe the two out-of-window rollouts that the benchmark defines, one in each temporal direction, both scored against real HILDA+ ground truth. In the \emph{backward} rollout the model is conditioned on the 1960 map and unrolled to reconstruct 1955--1959, a five-year reach before the 1960 training start under purely historical forcing. In the \emph{forward} rollout it is conditioned on the 2014 map and unrolled to 2015--2019; because the historical LUH2 forcing ends in 2015, the forward drivers switch to the future-scenario (SSP) tracks, making this a genuine test of the model under scenario forcing. Both windows lie entirely outside the training and testing period, so each is a true temporal extrapolation.

Table~\ref{tab:extrap} reports both directions. In the forward direction, the learned models retain most of their in-window skill: overall accuracy falls only $\sim$2--3 points relative to the 2001--2014 rollout, and every learned model stays far above the frozen-persistence change floor in both directions. More importantly, the autoregressive signature identified in Section~\ref{sec:res_rollout} reappears symmetrically outside the training window: the AR variants drift over the rollout (forward: AR U-Net $-3.7$, AR Swin-ViT $-4.5$ points; backward: $-4.4$ and $-5.4$ points over five years), while the non-AR variants stay flat ($<0.3$ points) in both directions. That the trade-off survives in both extrapolation regimes is evidence it is intrinsic to the autoregressive design, not an artifact of the particular 2001--2014 test window.

Two properties of the forward rollout are worth noting. First, the three SSP scenarios (SSP2-4.5, SSP3-7.0, SSP4-6.0) yield results identical to within $0.001$ on every metric---over a five-year horizon the scenario drivers have not yet diverged, so extrapolation cost is governed by distance from the training period, not by scenario choice; we therefore report SSP2-4.5 as representative. Second, the apparent parity of the AR models' out-of-window and in-window \emph{averages} should not be read as an extrapolation gain: the in-window average spans fourteen rollout years, over which AR error compounds, whereas each extrapolation window spans only five, so the AR average is simply dragged down less. Within each five-year window the AR drift is plainly present. These windows are deliberately short---five years each side, over which only $\sim$0.2\% of pixels change annually---so we frame them as a controlled near-window extrapolation test, with reconstruction toward 1850 and projection toward 2100 remaining the longer-horizon target of future work.

\begin{table}[ht!]
    \centering
    \caption{\textbf{Out-of-window extrapolation} (5-seed means, 5-year rollout averages). \emph{Backward}: 1955--1959, conditioned on 1960, historical forcing. \emph{Forward}: 2015--2019, conditioned on 2014, SSP2-4.5 forcing (SSP3-7.0 and SSP4-6.0 agree to within $0.001$). Both windows are strictly outside the 1960--2000 training period. 
    All learned models exceed the frozen-persistence change floor ($0.163$).
    }
    \label{tab:extrap}
    \small
    \begin{tabular}{lcccc}
        \toprule
        & \multicolumn{2}{c}{\textbf{Backward (1955--1959)}} & \multicolumn{2}{c}{\textbf{Forward (2015--2019)}} \\
        \cmidrule(lr){2-3}\cmidrule(lr){4-5}
        \textbf{Model} & \textbf{Overall Acc.} & \textbf{Change Acc.} & \textbf{Overall Acc.} & \textbf{Change Acc.} \\
        \midrule
        AR U-Net          & 0.8246 & 0.3207 & 0.8140 & 0.4212 \\
        non-AR U-Net      & 0.8425 & \textbf{0.3559} & 0.8259 & 0.3923 \\
        AR Swin-ViT       & 0.8130 & 0.3010 & 0.8032 & \textbf{0.4452} \\
        non-AR Swin-ViT   & \textbf{0.8583} & 0.3312 & \textbf{0.8303} & 0.4131 \\
        Simple CNN        & 0.8280 & 0.2987 & 0.8082 & 0.4188 \\
        Linear            & 0.8072 & 0.3361 & 0.7993 & 0.3663 \\
        \bottomrule
    \end{tabular}
\end{table}

\begin{figure}[ht!]
  \centering
  \includegraphics[width=0.92\linewidth]{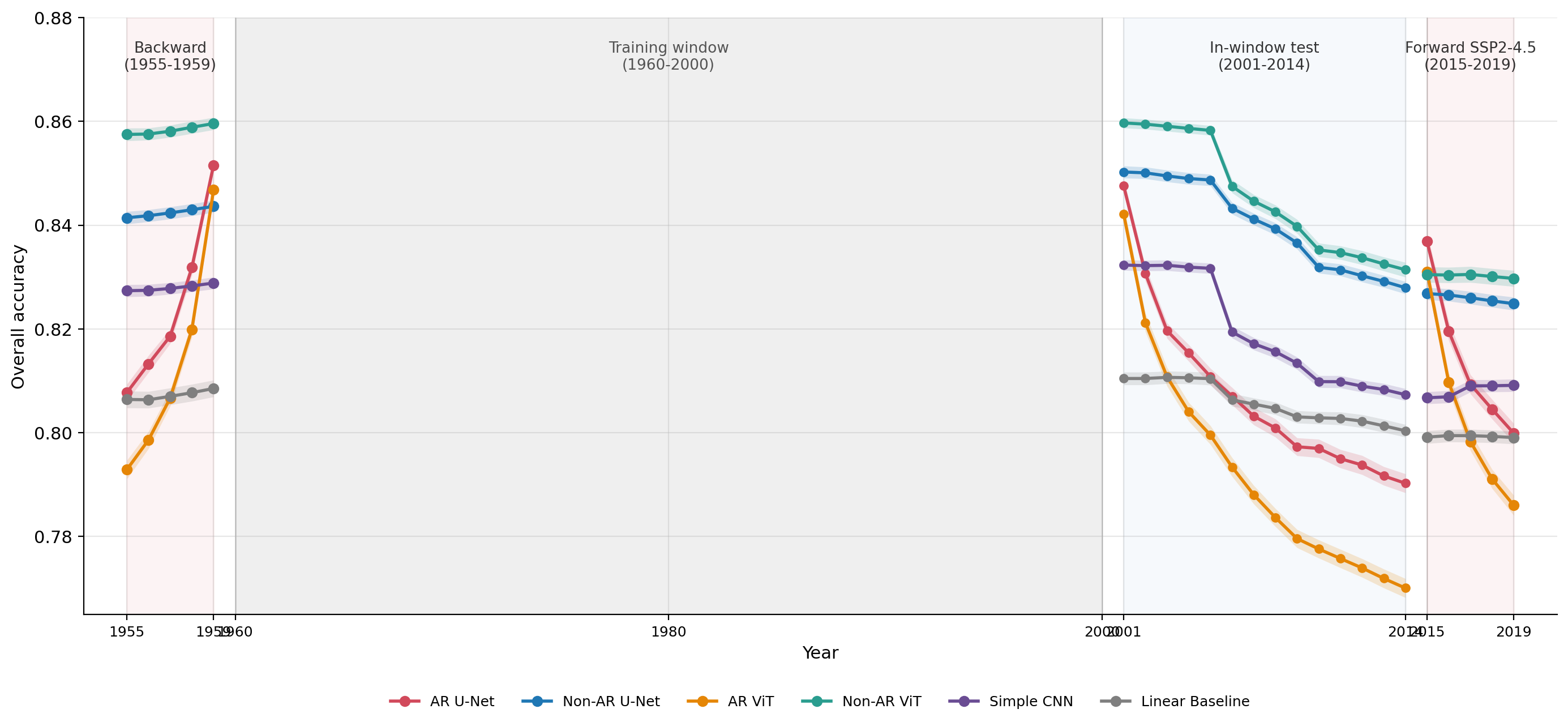}
  \caption{\textbf{Extrapolation in both temporal directions.} Per-year overall accuracy for the backward (1955--1959) and forward (2015--2019, SSP2-4.5) rollouts, framing the in-window 2001--2014 period. Autoregressive models drift away from the conditioning anchor in both directions; non-\hspace{0pt}autoregressive models remain flat.}
  \label{fig:extrap}
\end{figure}

\section{Discussion}
Our results expose a tension at the heart of autoregressive land-use modelling. The ablation identifies the previous-year map as by far the most informative input (its removal raises the loss by $16.7\%$), and conditioning on it gives the autoregressive models their advantage at detecting land-use change. Yet the very same dependence is a liability over long rollouts: feeding the model its own predictions back as the prior compounds error, costing roughly six accuracy points across fourteen years, whereas the non-\hspace{0pt}autoregressive models---which forgo the prior and predict each year independently---remain more stable and lead on overall accuracy, mIoU, and most per-class scores. Notably, this trade-off is governed by the autoregressive choice rather than the backbone: the backbone has a smaller, secondary effect, with the Swin-Transformer degrading somewhat faster than the convolutional U-Net over long horizons. For the multi-decadal product this framework ultimately targets, the implication is that a non-\hspace{0pt}autoregressive design---or a hybrid that periodically re-anchors to observations---is preferable for stability, while autoregression is better suited to short-range change detection.

The dual-regime evaluation is essential to reading these results correctly. Measured over all pixels, the trivial rolling-persistence baseline is effectively unbeatable ($0.99$ accuracy) because the land-use map barely changes from year to year; only the change-aware regime reveals that the learned models possess real skill ($0.35$--$0.39$ accuracy on changed pixels, against $0$--$0.16$ for persistence). Reporting both regimes is therefore necessary to avoid mistaking background dominance for predictive ability---a caution we build into the benchmark and recommend for the field. At the same time, the change-aware regime shows how much room remains: even the best models classify only about a third of the changed pixels correctly ($0.35$--$0.39$). A natural extension is therefore to treat reconstruction as a composed task: a dedicated change detector first identifies the pixels likely to transition in a given year, and a second model assigns the new class only to those pixels, leaving the rest unchanged. This separates the rare-event detection problem from the classification problem and directly targets the low accuracy on changed pixels. A dedicated analysis of where and how often the models predict change is the subject of follow-up work.

Supervision spans 1960--2000 and the primary evaluation is on 2001--2014, all within the satellite-era distribution of climate forcings and land-use trajectories. To probe behaviour beyond that window we evaluated the benchmark's two short extrapolation rollouts (Section~\ref{sec:res_extrap}): a backward reconstruction to 1955--1959 and a forward projection to 2015--2019 under SSP scenario forcing. Reassuringly, the learned models lose only $\sim$2--3 accuracy points relative to the in-window rollout in in the forward direction, and the autoregressive-versus-non-\hspace{0pt}autoregressive trade-off reproduces symmetrically outside the training window, indicating the trade-off is a property of the design rather than of the test period. These reaches are deliberately short---five years each side---and the near-invariance of the three SSP forcings over this horizon means the forward test stresses temporal distance more than scenario choice. Genuine multi-decadal extrapolation---reconstruction toward 1850 or projection toward 2100, where forcings leave the observed envelope and the SSP scenarios diverge---remains beyond what can be verified against ground truth and is the subject of follow-up work. Alongside this, rare and spatially fragmented classes (Urban, Grass/Shrubland) remain difficult for all models; and the autoregressive Water collapse reflects the absence of any open-water predictor in the coarse forcing, which leaves the class dependent on a prior that itself degrades under rollout.

\section{Conclusion}
We presented a deep-learning framework and the TerraNostra-Bench benchmark for high-resolution reconstruction of global land use, together with a dual-task, change-aware evaluation protocol. Comparing convolutional (U-Net) and transformer (Swin) backbones in autoregressive and non-\hspace{0pt}autoregressive schemes against persistence, linear, and convolutional baselines, we find that the autoregressive prior is simultaneously the most informative input and the principal source of long-horizon error, producing a stability--sensitivity trade-off that the choice of backbone does not override. Feature ablations explain this: the prior is \emph{substitutive} rather than additive. When it is present, removing all LUH2 or all static drivers leaves the loss essentially unchanged, whereas without it these drivers become essential (removing all LUH2 features raises the non-\hspace{0pt}autoregressive loss by $27\%$). Autoregressive models therefore learn to rely on a self-generated signal that accumulates error under rollout, while non-\hspace{0pt}autoregressive models remain anchored to external drivers. Future work will extend training with longer-rollout curricula to curb error accumulation and, building on the short backward and forward extrapolation tests reported here, apply the framework to the full multi-decadal historical reconstruction (toward 1850) and future projection (toward 2100) that motivates it.

\section*{Acknowledgements}

We acknowledge EuroHPC Joint Undertaking for awarding the project ID EHPC-DEV-2025D03-112, EHPC-DEV-2026D03-048, EHPC-DEV-2026D05-047, EHPC-BEN-2026B04-072,  EHPC-AIF-2025SC01-004, EHPC-AIF-2026SC02-008 access to MareNostrum5 at BSC, Spain.

Generative AI was used solely for polishing the language of this manuscript; it was not used for generating content, text, or figures, nor for performing any analysis. All AI-assisted text was reviewed and verified by the authors, who take full responsibility for the accuracy and integrity of the manuscript.

\section*{Data availability}

All input datasets used in this study are publicly available from their original providers:
HILDA+ global land-use change data \citep{winkler2020hilda+} (PANGAEA, \url{https://doi.pangaea.de/10.1594/PANGAEA.921846});
LUH2 land-use harmonization data \citep{hurtt2020harmonization} (\url{https://luh.umd.edu});
K\"oppen--Geiger climate classification maps \citep{beck2023high} (\url{https://doi.org/10.6084/m9.figshare.21789074});
ISIMIP2b population data \citep{frieler2017assessing} (ISIMIP repository, \url{https://data.isimip.org});
GEBCO 2024 elevation and bathymetry \citep{GEBCO2024} (\url{https://www.gebco.net});
and PNNL global soil composition data \citep{li2024global} (\url{https://zenodo.org/records/10523833}).
The TerraNostra-Bench benchmark (processed predictors, targets, spatial--temporal splits and evaluation protocol), trained model checkpoints, and training and evaluation code will be deposited at \url{https://zenodo.org/records/23040508} and will be made publicly available upon publication; they are available to editors and reviewers during peer review.

\section*{Conflict of interest}
The authors declare no competing interests.

\section*{Ethics statement}
This study did not involve human participants, human data or tissue, or animals, and therefore did not require ethical approval. The population data used are publicly available, aggregated gridded estimates and contain no personal or individually identifiable information.

\section*{Funding}
This work received funding from the European Union's Horizon Europe Framework Programme through the project CONCERTO (Grant Agreement 101185000), TerraDT (Grant Agreement 101187992) and ELLIOT (Grant Agreement 101214398). AM acknowledges his Grant JDC2023-051208-I, funded by MICIU/AEI/10.13039/501100011033 and, as appropriate, by “ERDF A way of making Europe”, by “ERDF/EU”, by the “European Union”, or by the “European Union NextGenerationEU/ PRTR”. AD and SM acknowledge their AI4S fellowship within the “Generación D” initiative by Red.es, Ministerio para la Transformación Digital y de la Función Pública, for talent attraction (C005/24-ED CV1), funded by NextGenerationEU through PRTR.


\section*{Author contributions}
\textbf{AM}: Conceptualization, Methodology, Software, Writing -- original draft,
Funding acquisition.
\textbf{MC}: Methodology, Software, Data curation, Writing -- review \& editing.
\textbf{SM}: Conceptualization, Methodology, Writing -- review \& editing, Funding acquisition.
\textbf{JVC}:  Development, Data curation, Writing.
\textbf{JVA}: Methodology, Software, Writing -- review \& editing.
\textbf{DGG}: Methodology, Software, Writing -- review \& editing, Funding acquisition.
\textbf{ET}: Writing -- review \& editing, Funding acquisition.
\textbf{MA}: Writing -- review \& editing, Funding acquisition.
\textbf{AD}: Conceptualization, Methodology, Writing -- review \& editing, Funding acquisition.
All authors reviewed and approved the final manuscript.

\clearpage
\appendix

\section{Spatial split and cross-validation}\label{app:split}

\paragraph{Macro-grid split.}
The global land surface is partitioned by binning the globe into a $40\times40$
macro-grid and assigning each cell \emph{in its entirety} to a single subset,
targeting a 70/15/15 train/validation/test ratio by land-pixel count;
Farthest Point Sampling then draws $256\times256$ patches
within each cell. Because no macro-cell contributes to more than one subset,
neighbouring pixels cannot leak across the partition, and the reported test
scores measure generalisation to spatially disjoint regions rather than
interpolation between adjacent training pixels.

\begin{figure}[ht!]
  \centering
  \includegraphics[width=\linewidth]{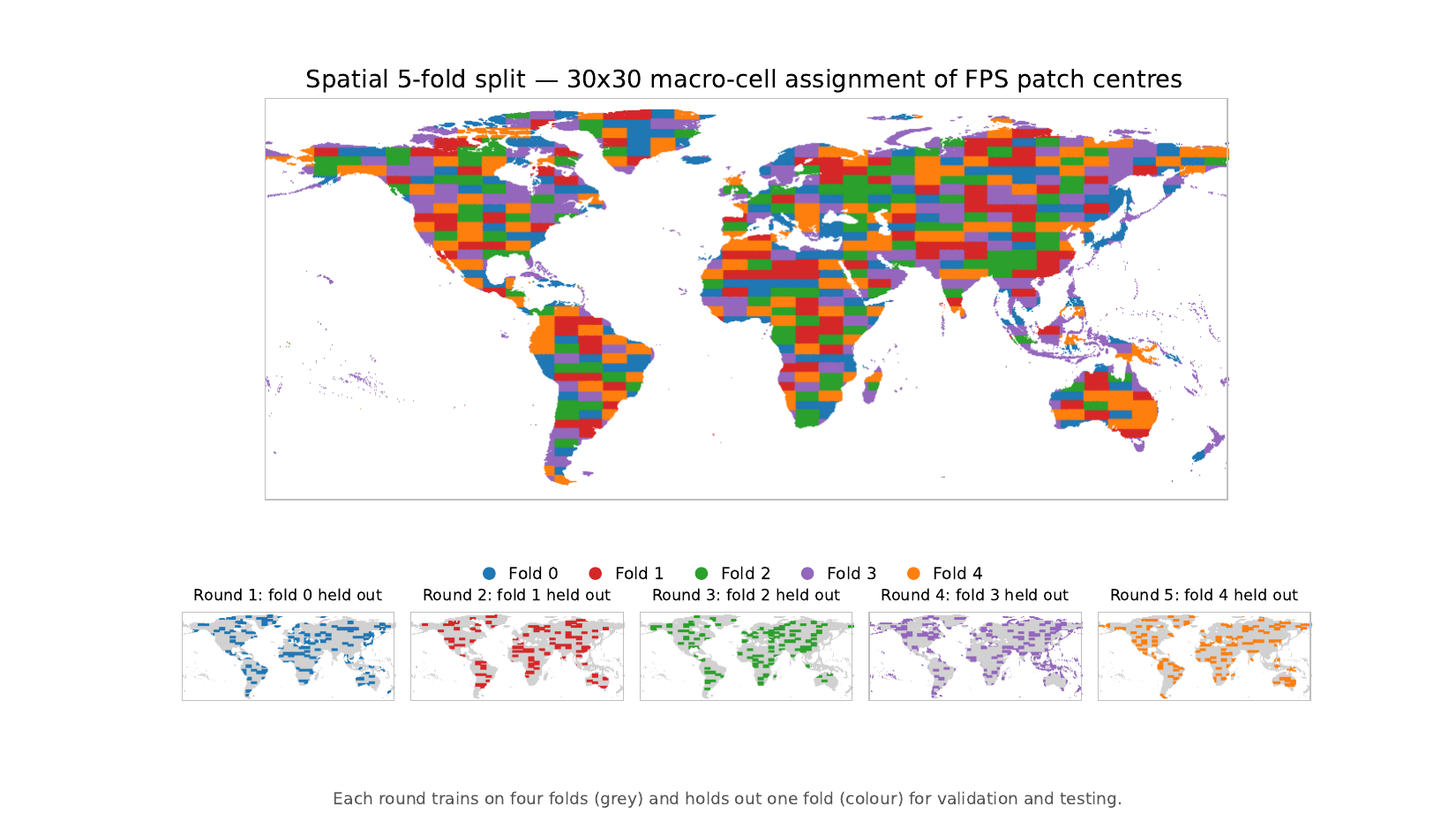}
  \caption{\textbf{Spatial five-fold split.} Top: FPS patch centres coloured by
  their exclusive $40\times40$ macro-cell fold assignment. Bottom: the five
  cross-validation rounds; each round trains on four folds (grey) and holds out
  one fold (colour) for testing. The canonical 70/15/15 split
  used in the main results is drawn from the same macro-cell machinery.}
  \label{fig:datasplit}
\end{figure}

\paragraph{Five-fold cross-validation.}
To verify that the reported scores are not an artifact of the particular
macro-cell assignment, we repeat the assignment with the same machinery into
five folds of ${\sim}20\%$ of land pixels each (90{,}000 patches per fold; the
assignment is drawn once and frozen). Figure~\ref{fig:datasplit} shows the fold
map and the five resulting cross-validation rounds: in round $i$, the model is
trained from scratch on the four remaining folds (${\sim}360$k patches) and
fold $i$ is held out, serving as
test region. Each round trains the non-\hspace{0pt}autoregressive U-Net at a matched
compute budget (best-validation checkpoint at epoch ${\sim}8$, aligned with the
epoch-9 reference sweep on the canonical split) and is evaluated with the same
14-year recurrent protocol as the main results.

\begin{table}[ht!]
  \centering
  \caption{\textbf{Split sensitivity of the non-AR U-Net.} Fourteen-year
  rollout means on the held-out fold of each cross-validation round, against
  the canonical-split reference (epoch-9 checkpoint; $\pm$ std over five
  evaluation seeds, i.e.\ pure sampling noise). Class-balanced metrics (macro
  F1, mean IoU) are insensitive to the split; overall accuracy varies with the
  regional class composition of the held-out area, as expected for spatial
  cross-validation.}
  \label{tab:foldsensitivity}
  \begin{tabular}{lccccc}
    \toprule
    & Overall acc. & Macro F1 & Mean IoU & Change acc. & Change F1 \\
    \midrule
    Fold 0                       & 0.793 & 0.688 & 0.562 & 0.429 & 0.281 \\
    Fold 1                       & 0.789 & 0.708 & 0.581 & 0.477 & 0.317 \\
    Fold 2                       & 0.764 & 0.712 & 0.582 & 0.452 & 0.311 \\
    Fold 3                  & 0.785 & 0.692 & 0.566 & 0.437 & 0.297 \\
    Fold 4                  & 0.757 & 0.688 & 0.558 & 0.434 & 0.285 \\
    \midrule
    Fold mean $\pm$ std     & $0.778 \pm 0.016$ & $0.698 \pm 0.011$ & $0.570 \pm 0.011$ & $0.446 \pm 0.019$ & $0.298 \pm 0.016$ \\
    Canonical split  & $0.846 \pm 0.001$ & $0.704 \pm 0.001$ & $0.582 \pm 0.001$ & $0.440 \pm 0.001$ & $0.259 \pm 0.001$ \\
    \bottomrule
  \end{tabular}

\end{table}

\begin{table}[ht!]
  \centering
  \caption{\textbf{Split sensitivity of the AR U-Net.} Fourteen-year rollout
  means on the held-out fold of each cross-validation round (best-validation
  checkpoints at epochs 2--3 under a fixed training budget), against the
  canonical-split model evaluated at matched maturity (epochs 2 and 3) and at
  its best-validation checkpoint (epoch 4; the setup reported in the main
  text). At matched budget the canonical scores are consistent with the fold
  distribution, but the fold spread is four times larger than for the non-AR
  model (Table~\ref{tab:foldsensitivity}) and orders by the land-use-change
  fraction of the held-out regions (folds~2 and~4, ${\sim}46\%$ change pixels,
  score lowest) --- the HILDA prior that feeds the autoregressive design is
  less informative where landscapes change.}
  \label{tab:foldsensitivity_ar}
  \begin{tabular}{lccccc}
    \toprule
    & Overall acc. & Macro F1 & Mean IoU & Change acc. & Change F1 \\
    \midrule
    Fold 0                        & 0.738 & 0.569 & 0.458 & 0.420 & 0.269 \\
    Fold 1                        & 0.777 & 0.667 & 0.545 & 0.457 & 0.267 \\
    Fold 2                        & 0.643 & 0.577 & 0.448 & 0.423 & 0.240 \\
    Fold 3                        & 0.746 & 0.601 & 0.484 & 0.413 & 0.254 \\
    Fold 4                        & 0.614 & 0.553 & 0.424 & 0.419 & 0.262 \\
    \midrule
    Fold mean $\pm$ std           & $0.704 \pm 0.071$ & $0.593 \pm 0.045$ & $0.472 \pm 0.046$ & $0.426 \pm 0.018$ & $0.258 \pm 0.012$ \\
    Canonical split      & 0.719 & 0.527 & 0.429 & 0.427 & 0.278 \\
    \bottomrule
  \end{tabular}
\end{table}

\paragraph{Interpretation.}
For the non-AR U-Net, the class-balanced quality metrics match the canonical
split closely: macro F1 of $0.698 \pm 0.011$ across folds against $0.704$ on
the canonical split, and mean IoU of $0.570 \pm 0.011$ against $0.582$.
Change-pixel metrics are on par with or slightly above the canonical values.
Overall accuracy is lower and more variable across folds ($0.757$--$0.793$
vs.\ $0.846$); since it is dominated by the majority classes, it reflects the
class composition of the specific fifth of the globe held out in each round
rather than model quality --- consistent with the class-balanced metrics being
tied. The AR U-Net behaves differently: at matched training maturity its
canonical scores are consistent with the fold distribution (fold mean $0.704
\pm 0.071$ vs.\ $0.719$/$0.792$ at canonical epochs 2/3), but its fold spread
is four times larger than the non-AR model's and orders by the
land-use-change fraction of the held-out regions (folds 2 and 4,
${\sim}46\%$ change pixels vs.\ $21\%$ on the canonical test regions, score
lowest). This is the expected signature of the autoregressive design: the
HILDA prior it feeds on is less informative where landscapes change, so its
performance depends on where it is tested --- and the canonical test
regions, being atypically static, sit at the favourable end of the fold
range.

\section{Qualitative rollout comparison}\label{app:drift}

\begin{figure}[ht!]
  \centering
  
  \includegraphics[width=0.82\linewidth]{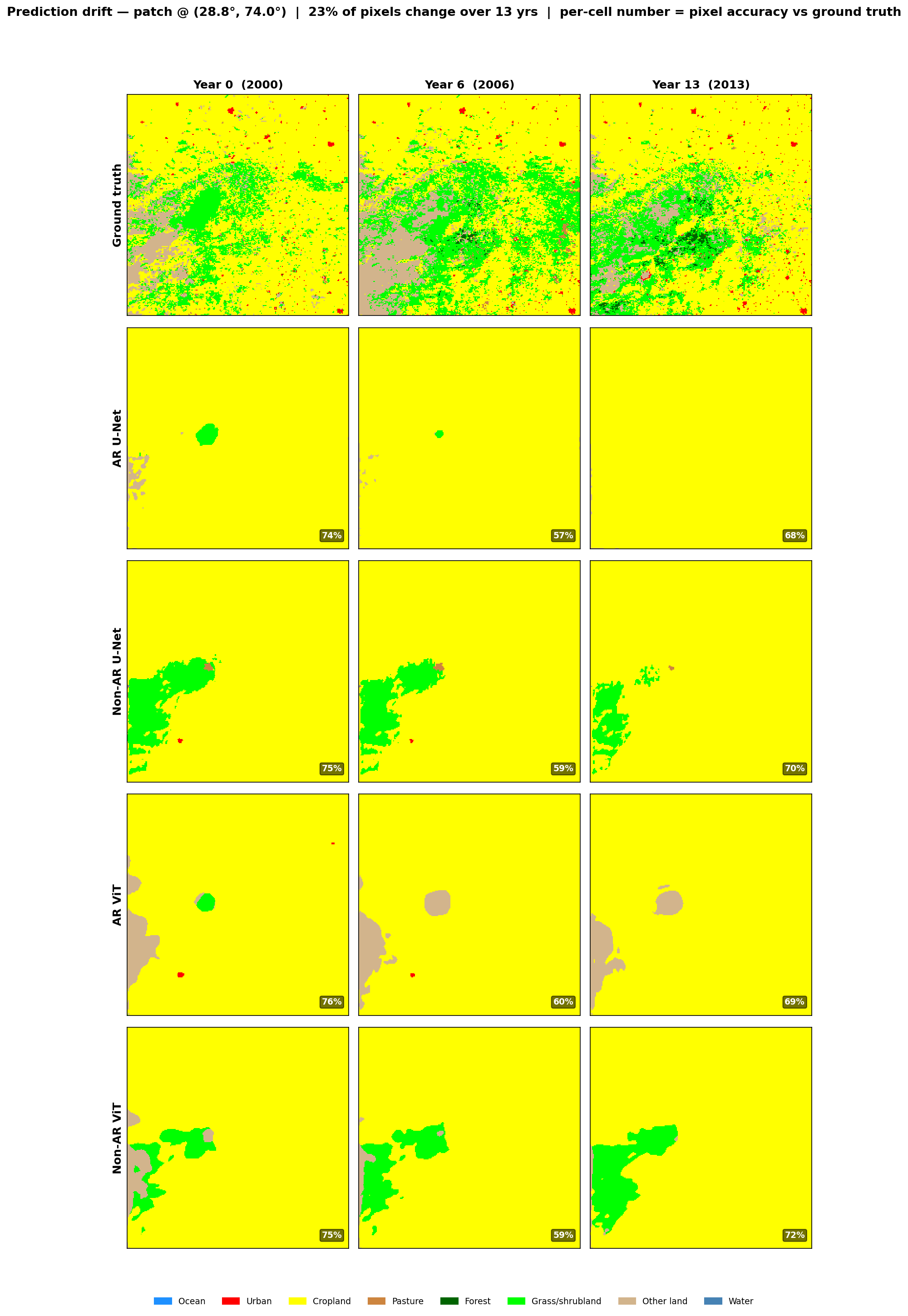}
  \caption{\textbf{High-change patch, northwest India} ($28.8^\circ$N, $74.0^\circ$E; 23\% of pixels change class between 2000 and 2013). Rows: ground truth and the four models; columns: rollout years 0 (2000), 6 (2006) and 13 (2013). The number in each panel is its pixel accuracy against ground truth. All four models stay within a few points of each other throughout (68--72\% at year~13), dip at year~6 and partly recover; all smooth the grass/shrubland mosaic into cropland.}
  \label{fig:drift_main}
\end{figure}

\begin{figure}[ht!]
  \centering
  \includegraphics[width=0.78\linewidth]{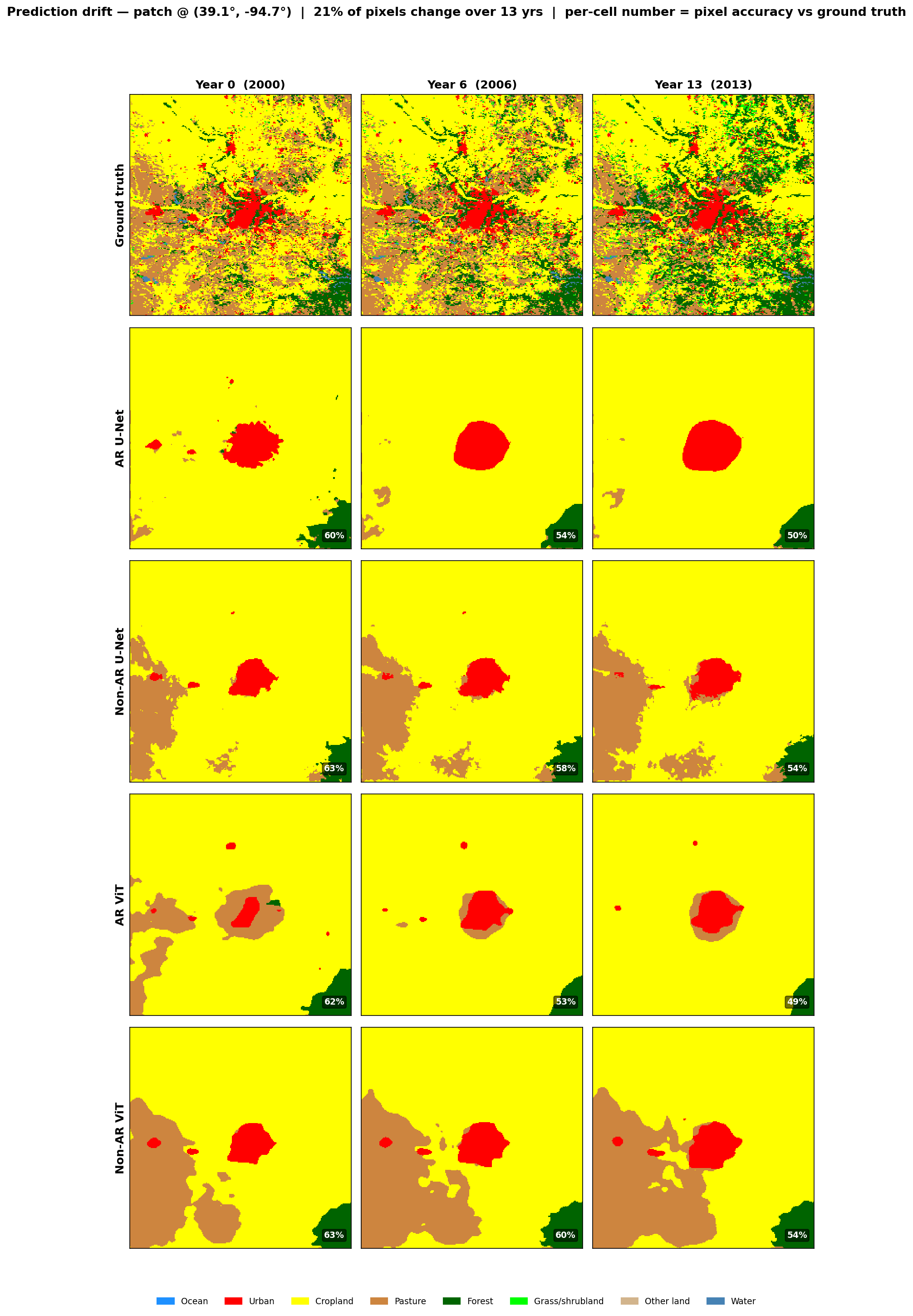}
  \caption{\textbf{High-change patch, Kansas City, USA} ($39.1^\circ$N, $94.7^\circ$W; 21\% change). Layout as in Figure~\ref{fig:drift_main}. Accuracy falls for all models, slightly more for the autoregressive variants (AR U-Net $60\!\rightarrow\!50\%$, AR Swin-ViT $62\!\rightarrow\!49\%$) than for the non-\hspace{0pt}autoregressive ones ($63\!\rightarrow\!54\%$ for both). All models capture the urban core but smooth the surrounding pasture--cropland mosaic.}
  \label{fig:drift_high2}
\end{figure}

\begin{figure}[ht!]
  \centering
  \includegraphics[width=0.78\linewidth]{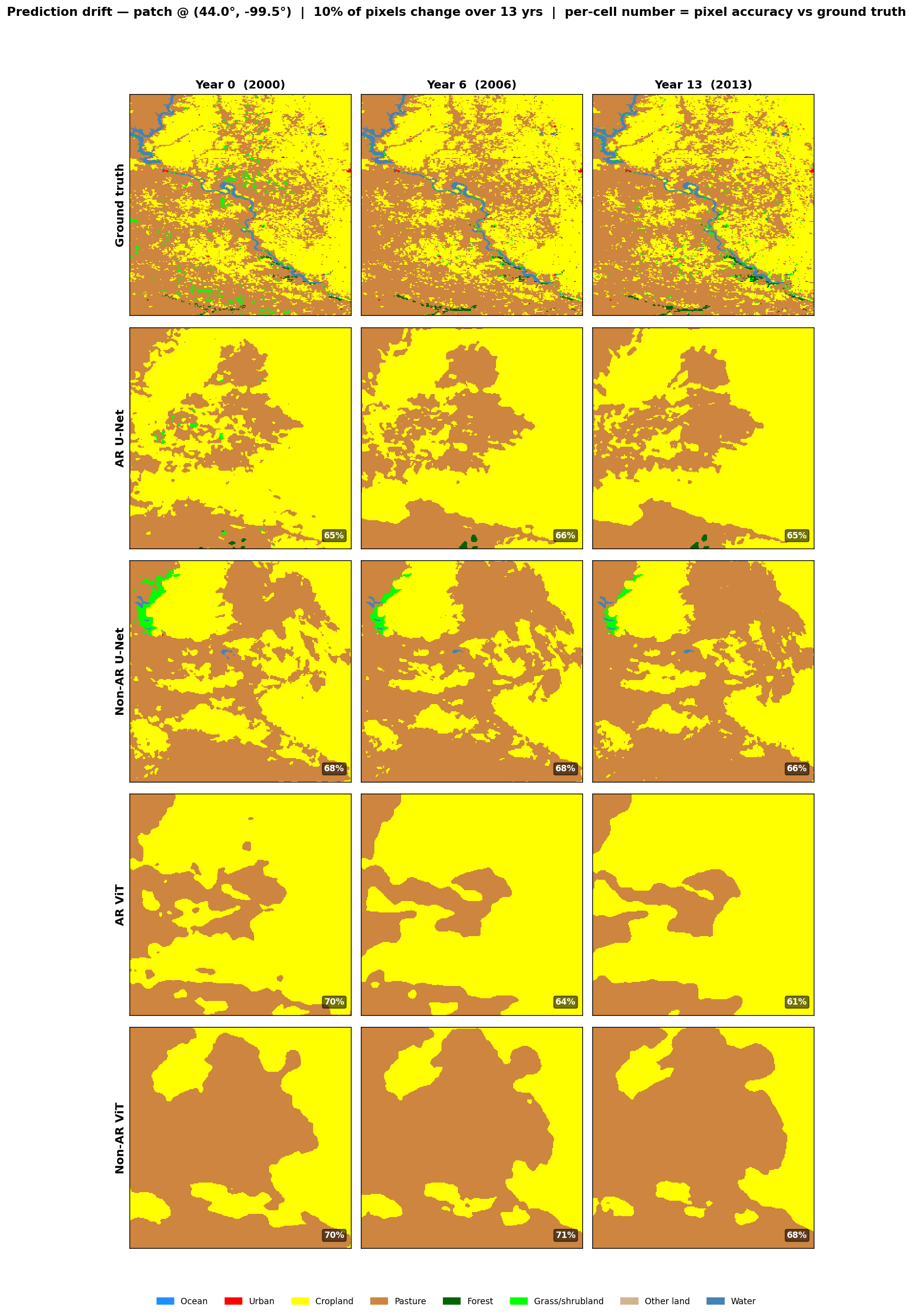}
  \caption{\textbf{Moderate-change patch, US Great Plains} ($44.0^\circ$N, $99.5^\circ$W; 10\% change). Layout as in Figure~\ref{fig:drift_main}. The non-\hspace{0pt}autoregressive models remain nearly flat (U-Net $68\!\rightarrow\!66\%$, Swin-ViT $70\!\rightarrow\!68\%$), as does the AR U-Net ($65\%$ throughout), whereas the AR Swin-ViT declines from $70\%$ to $61\%$. All models reproduce the large-scale pasture--cropland pattern but miss the river and its riparian vegetation.}
  \label{fig:drift_mid1}
\end{figure}

\begin{figure}[ht!]
  \centering
  \includegraphics[width=0.78\linewidth]{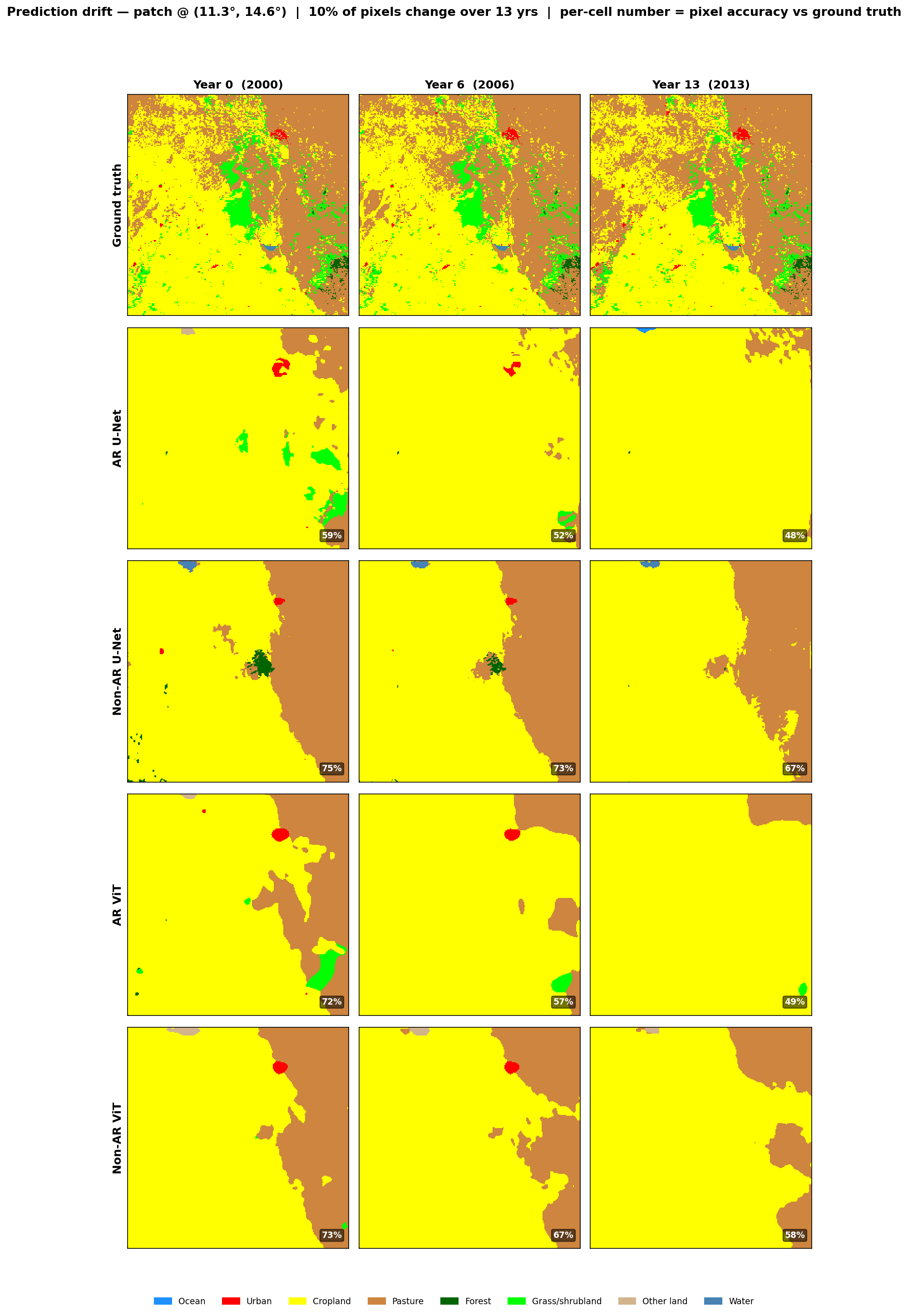}
  \caption{\textbf{Moderate-change patch, northern Cameroon} ($11.3^\circ$N, $14.6^\circ$E; 10\% change). Layout as in Figure~\ref{fig:drift_main}. The clearest example of autoregressive drift: the AR U-Net falls from $59\%$ to $48\%$ and the AR Swin-ViT from $72\%$ to $49\%$ as both progressively overwrite the eastern pasture with cropland, whereas the non-\hspace{0pt}autoregressive U-Net stays largely stable ($73\!\rightarrow\!67\%$) and the non-\hspace{0pt}autoregressive Swin-ViT declines less ($73\!\rightarrow\!58\%$).}
  \label{fig:drift_mid2}
\end{figure}

Figures~\ref{fig:drift_main}--\ref{fig:drift_mid2} compare the four models on two high-change and two moderate-change test patches. Per-patch behaviour varies: autoregressive drift is pronounced in some regions (Figure~\ref{fig:drift_mid2}) and largely absent in others (Figure~\ref{fig:drift_main}), which is why our conclusions rest on the aggregate rollout statistics in Section~\ref{sec:res_rollout} rather than on individual patches.

\textbf{Patch selection.} Patches were drawn from the held-out spatial test
split using a model-independent change metric: the fraction of pixels whose
HILDA+ class at rollout year~13 (2013) differs from year~0 (2000). High-change
patches were the five patches with the largest such change in a random sample
of 150 test patches (17.9--25.6\% change); moderate-change patches were drawn
from a random sample of 250 test patches, restricted to those at least 80\%
land with 2000$\rightarrow$2013 change between 5 and 15\%, taking the five
closest to the 10\% midpoint ($\approx$9.8--10.2\% change). The same patch and
inputs are given to all four models, which are then run for the full 14-year
recurrent rollout and displayed at years 0, 6, and 13. Selection samples a
random subset rather than ranking the entire test set, so the patches are
representative rather than globally extreme; the 80\% land filter was applied
to the moderate-change set only, as high-change patches are intrinsically
land-dominated (the ocean class is static and cannot accrue change).

\bibliographystyle{plainnat}
\bibliography{refs}            

\end{document}